\documentclass[a4paper,12pt, epsfig]{article}
\usepackage{epsfig}
\usepackage{amssymb}
\usepackage{amsfonts}

\newskip\humongous \humongous=0pt plus 1000pt minus 1000pt

\newif\ifdtup

\catcode`\@=11

\@addtoreset{equation}{section}
\def\theequation{\thesection.\arabic{equation}}

\def\@normalsize{\@setsize\normalsize{15pt}\xiipt\@xiipt
\abovedisplayskip 14pt plus3pt minus3pt%
\belowdisplayskip \abovedisplayskip
\abovedisplayshortskip \z@ plus3pt%
\belowdisplayshortskip 7pt plus3.5pt minus0pt}

\def\small{\@setsize\small{13.6pt}\xipt\@xipt
\abovedisplayskip 13pt plus3pt minus3pt%
\belowdisplayskip \abovedisplayskip
\abovedisplayshortskip \z@ plus3pt%
\belowdisplayshortskip 7pt plus3.5pt minus0pt
\def\@listi{\parsep 4.5pt plus 2pt minus 1pt
     \itemsep \parsep
     \topsep 9pt plus 3pt minus 3pt}}

\relax

\catcode`@=12

\catcode`\@=11

\def\section{\@startsection{section}{1}{\z@}{3.5ex plus 1ex minus
   .2ex}{2.3ex plus .2ex}{\large\bf}}

\def\thesection{\arabic{section}}
\def\thesubsection{\arabic{section}.\arabic{subsection}}

\def\appendix{\setcounter{section}{0}
 \def\thesection{Appendix \Alph{section}}
 \def\thesubsection{\Alph{section}.\arabic{subsection}}
 \def\theequation{\Alph{section}.\arabic{equation}}}

\def\SymBoxes#1#2#3#4{\newdimen\un@t \un@t#3%
\raisebox{#1}{\rule{#2\un@t}{#4}\hskip-#2\un@t
\@tempdimb\un@t \advance\@tempdimb by-#4\@tempcntb#2\relax%
\@whilenum{\@tempcntb>0}\do{
\rule{#4}{\un@t}\hskip\@tempdimb \advance\@tempcntb by\m@ne}%
\hskip-#2\un@t \rule[\un@t]{#2\un@t}{#4}%
\rule[\un@t]{#4}{#4}\hskip-#4
\rule{#4}{\un@t}}\hskip-#4}                

\begin{document}

\newcommand{\beq}{\begin{equation}}
\newcommand{\eeq}{\end{equation}}
\newcommand{\bea}{\begin{eqnarray}}
\newcommand{\eea}{\end{eqnarray}}
\newcommand{\beas}{\begin{eqnarray*}}
\newcommand{\eeas}{\end{eqnarray*}}
\newcommand{\defi}{\stackrel{\rm def}{=}}
\newcommand{\non}{\nonumber}
\newcommand{\bquo}{\begin{quote}}
\newcommand{\enqu}{\end{quote}}
\renewcommand{\(}{\begin{equation}}
\renewcommand{\)}{\end{equation}}
\def\de{\partial}
\def\Tr{ \hbox{\rm Tr}}
\def\H{ \hbox{\rm H}}
\def\HE{ \hbox{$\rm H^{even}$}}
\def\HO{ \hbox{$\rm H^{odd}$}}
\def\K{ \hbox{\rm K}}
\def\Im{ \hbox{\rm Im}}
\def\Ker{ \hbox{\rm Ker}}
\def\const{\hbox {\rm const.}}
\def\o{\over}
\def\im{\hbox{\rm Im}}
\def\re{\hbox{\rm Re}}
\def\bra{\langle}\def\ket{\rangle}
\def\Arg{\hbox {\rm Arg}}
\def\Re{\hbox {\rm Re}}
\def\Im{\hbox {\rm Im}}
\def\exo{\hbox {\rm exp}}
\def\diag{\hbox{\rm diag}}
\def\longvert{{\rule[-2mm]{0.1mm}{7mm}}\,}
\def\a{\alpha}
\def\dag{{}^{\dagger}}
\def\tq{{\widetilde q}}
\def\p{{}^{\prime}}
\def\W{W}
\def\N{{\cal N}}
\def\hsp{,\hspace{.7cm}}
\newcommand{\C}{\ensuremath{\mathbb C}}
\newcommand{\Z}{\ensuremath{\mathbb Z}}
\newcommand{\R}{\ensuremath{\mathbb R}}
\newcommand{\rp}{\ensuremath{\mathbb {RP}}}
\newcommand{\cp}{\ensuremath{\mathbb {CP}}}
\newcommand{\vac}{\ensuremath{|0\rangle}}
\newcommand{\vact}{\ensuremath{|00\rangle}                    }
\newcommand{\oc}{\ensuremath{\overline{c}}}
\begin{titlepage}
\begin{flushright}
ULB-TH/06-17\\
hep-th/0607045\\
\end{flushright}
\bigskip
\def\thefootnote{\fnsymbol{footnote}}

\begin{center}
{\large {\bf
Can D-branes Wrap Nonrepresentable Cycles?
 } }
\end{center}

\bigskip
\begin{center}
{\large  Jarah EVSLIN\footnote{\texttt{ jevslin@ulb.ac.be}}$^1$ 
and Hisham SATI\footnote{\texttt{hisham.sati@adelaide.edu.au}}$^{2,3,4}$
 }
\end{center}

\renewcommand{\thefootnote}{\arabic{footnote}}

\begin{center}
\vspace{1em}
{\em  { $^1$ International Solvay Institutes,\\
Physique Th\'eorique et Math\'ematique,\\
Universit\'e Libre
de Bruxelles, \\ C.P. 231, B-1050, Bruxelles, Belgium\\
\vspace{0.2cm}

$^2$ Department of Mathematics,
Yale University\\
New Haven, CT 06520, USA\\
\vspace{0.2cm}

$^3$Department of Pure Mathematics, University of Adelaide\\
       Adelaide, SA 5005,
     Australia}\\
\vspace{0.2cm}
$^4$The Erwin Schr\"odinger International\\ 
Institute for Mathematical Physics,\\ 
Boltzmanngasse 9, A-1090 Wien, 
Austria}\\
\end{center}

\noindent
\begin{center} {\bf Abstract} \end{center}
Sometimes a homology cycle of a nonsingular compactification manifold 
cannot be represented by a nonsingular submanifold.  We want to know 
whether such nonrepresentable cycles can be wrapped by D-branes.  A 
brane wrapping a representable cycle carries a K-theory charge if and 
only if its Freed-Witten anomaly vanishes.  However some K-theory 
charges are only carried by branes that wrap nonrepresentable cycles.  We provide two examples of Freed-Witten anomaly-free D6-branes wrapping nonrepresentable cycles in the presence of a trivial NS 3-form flux.  The first occurs in type IIA string theory compactified on the 
Sp(2) group manifold and the second in IIA on a product of lens spaces.  
We find that the first D6-brane carries a K-theory charge while the second does not.


\vfill

\end{titlepage}
\bigskip

\hfill{}
\bigskip

\setcounter{footnote}{0}
\section{Introduction}

\noindent
D-branes are not classified by homology.  For example in type II string theory on a $spin$ spacetime with a topologically trivial NSNS 3-form flux, Freed and Witten \cite{FW} have shown that 
D-branes can only wrap $spin^c$ submanifolds.  Maldacena, Moore and Seiberg (MMS) 
\cite{MMS} demonstrated that even in the SU(3) WZW model some homology classes of the 
spacetime are not represented by $spin^c$ manifolds and so cannot be wrapped by D-branes. 
With such inconsistent D-branes removed from the spectra, MMS demonstrated that the conserved brane 
charges in this example are classified not by homology but by twisted K-theory, in 
line with the conjectures of Refs.~\cite{MM,WittenK, BM}.

In general the Freed-Witten (FW) anomaly is a necessary but not a sufficient condition for 
the homology class of a D-brane to lift to a twisted K-theory class.  We will argue that, when the NS 3-form $H$ is topologically trivial, the FW anomaly is a necessary and sufficient condition for all D$p$-branes except for D6-branes.  D6-branes are special because they wrap 7-dimensional cycles in a 10-dimensional spacetime.  Ren\'e Thom, in work that won him the 1958 Fields Medal, demonstrated that this is the lowest-dimensional case in which a homology cycle may not be representable by a nonsingular submanifold.  This leads us to the question: {\it{Can D-branes wrap nonrepresentable cycles?}}  To answer this question definitively one should look at the worldsheet theory of fundamental strings, impose boundary conditions corresponding to a singular representative of the cycle and check for inconsistencies, such as a failure of BRST invariance.  In the present note we will use a less reliable method.  We will check to see whether branes wrapping nonrepresentable cycles carry K-theory charges.  We will find that the answer is {\it{yes}} for some cycles and {\textit{no}} for others. 

Although one can show that FW anomaly-free D6-branes that fail to carry a K-theory charge always wrap nonrepresentable cycles, the converse does not always hold.  Some K-theory charges are only carried by nonrepresentable branes.  To demonstrate this, we will recall an example of a D6-brane on a nonrepresentable cycle from Ref.~\cite{BHK} and show that it does carry a K-theory charge\footnote{The generalized D-branes of Ref.~\cite{Bryan} may, implicitly, wrap nonrepresentable cycles.}.  Thus the type IIA version \cite{Petr} of the Sen conjecture \cite{Sen} implies that D-branes may wrap certain nonrepresentable cycles.

An example\footnote{We have been informed that the first demonstration of the nonrepresentability of this cycle is in version one of Ref.~\cite{BHK}, which is also available on the arXiv at the same URL as the current version. However this reference does not address the issue of whether a brane wrapping this cycle carries a K-theory charge, which is the focus of the current note.} of a FW anomaly-free D6-brane that does not carry a K-theory charge was presented in Refs.~\cite{BRST,Hisham}.  We will argue that 
the wrapped 7-cycle is not represented by any nonsingular submanifold, which in particular implies that its singularity cannot be removed by deformations or blowups.  We find that the singular locus is homologous to a two-torus.  

In Sec.~\ref{background} we explain the relation between homology and K-theory charges.  
We review an algorithm for calculating K-theory from homology, and in particular we 
find the obstructions to lifting a homology class to a K-theory class.  In 
Sec.~\ref{illustrations} we use a result of Ren\'e Thom to argue that all of 
these obstructions are summarized by the Freed-Witten anomaly together with a necessary but not sufficient condition for the representability of the wrapped homology cycle by a nonsingular submanifold.  Then in Sec.~\ref{ex} we 
will present two examples of the second kind of obstruction, a D6-brane wrapping a 
nonrepresentable 7-cycle in the group manifold Sp(2) and in a product of 
lens spaces. We end with some discussion 
in Sec.~\ref{dis}. 


\section{The Atiyah-Hirzebruch Spectral Sequence}
\label{background}

A D-brane that wraps a nontrivial cycle carries a charge that corresponds to the homology
class of the cycle.  Diaconescu, Moore and Witten (DMW) ~\cite{DMW} have shown that not 
all of these charges are conserved, instead there are dynamical processes in which branes 
wrapping nontrivial cycles can decay.  In addition, in Ref.~\cite{FW} the authors have 
found that certain cycles cannot be wrapped by single branes. They argued that any brane 
wrapping such a cycle would be anomalous, and in fact evidence was presented in \cite{DMW} 
that their contributions to the partition function cancel.  Thus to compute the partition
function it suffices to restrict one's attention to equivalence classes of anomaly-free 
branes.  In other words, D-branes are classified by a quotient of a subset of homology.  

MMS have argued that this quotient of a subset is precisely twisted K-theory.  They used 
a mathematical algorithm known as the Atiyah-Hirzebruch spectral sequence (AHSS) to 
determine which homology classes lift to K-theory classes, that is, to determine which 
D-branes are unstable and which are not allowed.   While in their examples the anomalous 
branes suffered from the Freed-Witten (FW) anomaly, in general the AHSS construction 
eliminates some branes that are FW anomaly-free.  This leads to the question of whether 
the branes that are eliminated by the AHSS construction, but not by the FW anomaly, are 
allowed in the physical theory.  If such branes are allowed, they would provide counterexamples to the 
K-theory classification program and to the IIA version of the Sen conjecture.  On the other hand, if such branes are not allowed, 
they would be examples of a new anomaly.  In the 
present note we will adopt the more modest goal of providing a characterization of these branes.

The AHSS consists of a series of differential operators $d_{2p-1}$, $p\geq 2$ which map 
elements of the $q$th integral cohomology to elements of the 
$(q+2p-1)$th cohomology
\beq
d_{2p-1}:\H^q\longrightarrow\H^{q+2p-1}\hsp p=2,3,\cdots .
\eeq
In general the differential $d_{2p-1}$ consists of cohomology operations on the free
parts of the integral cohomology and also on cyclic subgroups of prime order less than 
or equal to $p$.  In particular, in the case of untwisted K-theory and when $p$ is prime, 
$d_{2p-1}$ contains a primary cohomology operation known as the first Milnor primitive
\footnote{Higher Milnor primitives appear in the differentials of `higher' 
generalized cohomology theories, for example $Q_2$ appears in Morava 
K-theory and elliptic cohomology \cite{KS1}.}  
$Q_1$ whose image is a $p$ torsion class $\Z_p$, and also it may contain secondary 
operations whose images are torsion at lower primes.    A secondary cohomology operation 
is an operation that is not defined on the entire cohomology, but is defined on the 
kernels of the preceding differentials.

We will use Poincar\'e duality to identify a D$(9-q)$-brane wrapping a $(10-q)$-cycle 
in the integral homology group $\H_{10-q}$ with its dual cocycle in $\H^q$, which in 
terms of supergravity fields corresponds to the Ramond-Ramond source $dG_{q-1}$.  The 
homology class of a D-brane wrapping the cycle $N_{10-q}$ lifts to a twisted K-theory 
class if and only if its dual cohomology class PD($N_{10-q})$ is in the kernel of all of the differentials
\beq
d_{2p-1} (\textup{PD}(N_{10-q}))=0 \hspace{.3cm} \textup{\ for\ all\ }p.
\eeq

For example, the first nontrivial differential contains a primary operation at prime 2 
and can be explicitly written
\beq
d_3 x=Q_1x+H\cup x=Sq^3x+H\cup x
\eeq
where $H$ is the NSNS 3-class and $\cup$ is the cup product, the integral version of the wedge product.  
The Milnor primitive $Q_1$ at prime 2 is often denoted $Sq^3$ and is called a Steenrod square or more precisely square 3.  $Sq^3$, like the cup product with $H$, increases the degree of a cohomology class by three.  DMW have explained that 
if $d_3$ does not annihilate the class of a brane then the brane suffers from an FW 
anomaly.  The converse is not true since some FW anomalous branes are annihilated 
by $d_3$.  MMS have found an example of this phenomenon in the SU(3) 
Wess-Zumino-Witten WZW model, and in that case the offending class was not in the 
kernel of $d_5$ and so, as expected, did not lift to twisted K-theory.  

More concretely, consider a brane with worldvolume $N$ in the spacetime $M$.  Let 
$i:N\hookrightarrow M$ be the inclusion map of the brane into the spacetime.  Then the 
FW anomaly is \cite{FW}
\beq
W_3+H=0
\eeq
where $W_3$ is the third integral Stiefel-Whitney class of the normal bundle of $N$ in $M$ and 
$H$ is the pullback of the NSNS 3-form to the brane worldvolume $N$.  The pushforward of 
$W_3+H$ to the spacetime $M$ is 
\beq
i_*(W_3+H)=Sq^3({\textup{PD}}(N))+H\cup\textup{PD}(N)=d_3(\textup{PD}(N)).
\eeq
In the aforementioned SU(3) example $W_3+H$ is nontrivial but it is in the kernel of the 
pushforward.  This example suggests that the role of the secondary operations is to pick 
up the anomalies that were in the kernel of the pushforward.  In particular one may 
conjecture that the secondary operations do not imply the existence of any new anomalies, for example all of the two-torsion operations encode the FW anomaly.


\section{Nonrepresentable Cycles}
\label{illustrations}
While the mod 2 Milnor primitive $Q_1=-Sq^3$ captures the FW anomaly, the mod 3 Milnor 
primitive $Q_1$ is insensitive to 2-torsion characteristic classes like the third 
Stiefel-Whitney class and so it describes a different anomaly.  Our next goal will be to 
characterize the worldvolumes of the branes suffering from this new anomaly.  We will 
restrict our attention to the case in which the NS $H$ flux is topologically trivial.  By this we mean not only that
$H$ is exact as a differential form but further that it represents the trivial class of the full integral cohomology.  

Imagine that the cycle $N$ wrapped by our D-brane is a nonsingular manifold with a $spin^c$ 
normal bundle, embedded as usual in a $spin$ spacetime.  Consider a trivial, rank 
one vector bundle on our D-brane.  This defines a nontrivial class in the K-theory of 
$N$.  As the normal bundle to $N$ is $spin^c$, we may push this class forward into the 
K-theory with compact support on a tubular neighborhood of $N$ in $M$.  We may then push 
this class forward yet again into the K-theory of $M$ without obstruction, and we will 
obtain the (possibly trivial) K-class which is the K-theory lift of the cohomology Poincar\'e dual of $N$.  
Thus the cohomology Poincar\'e dual of $N$ lifts to a K-class whenever $N$ is a nonsingular 
manifold with a $spin^c$ normal bundle.  The $spin^c$ normal bundle condition is physically just the 
condition that the brane not have a Freed-Witten anomaly.  The AHSS indicates that there 
must be other obstructions arising at other primes, but the existence of the above pushforward construction of the K-class suggests
that if $N$ is nonsingular then there is no other obstruction.  Thus the $Q_1$'s at higher 
primes, which appear in the higher differentials, may only be obstructions to the 
existence of a nonsingular manifold representing the homology class of $N$. 

This has been known in the mathematics literature for more than half a century.  Ren\'e 
Thom proved \cite{Thom} that any cohomology operation at an odd prime, at any dimension 
not equal to zero modulo four, annihilates all cohomology classes which are dual to 
homology classes that are representable by nonsingular manifolds.  The primes greater than 
two are all odd and the $Q_1$'s are all of odd degree, which are not equal to zero modulo 
four.  Therefore Thom's theorem implies that $Q_1$ at every prime greater than two 
annihilates all cohomology classes dual to cycles that can be represented by nonsingular
submanifolds.  

In particular, up to secondary operations, a D-brane which does not lift to a K-theory class is not in the kernel of 
$Q_1$ at some prime.  If the D-brane does not have an FW anomaly, then it is in the 
kernel of $Q_1$ at prime 2, and so it must not be in the kernel of $Q_1$ at some odd 
prime.  Therefore its homology class is not representable by a nonsingular manifold.

In critical superstring theories the spacetime is a 10-dimensional manifold.  
The lowest dimensional $Q_1$ that measures an obstruction to the representability of 
a homology class occurs at prime 3.  It is $Q_1=-\beta P^1_3$, where $\beta$ is the 
Bockstein homomorphism which raises the cohomology degree by one, 
\(
\beta~:~H^{2j+1}(X,\Z_3) \longrightarrow H^{2j+2}(X,\Z_3),
\)
arising from the exact sequence of coefficients
\(
0 \longrightarrow \Z_3 \longrightarrow \Z_{9} \longrightarrow \Z_3
\longrightarrow 0.
\)
$P^1_3$ is the first Steenrod power 
operation at the prime $3$
\(
P^1_3~:~ H^k(X,\Z_3) \longrightarrow H^{k+2(3-1)}(X,\Z_3).
\)

In particular $Q_1$ annihilates any cocycle of degree less than 3, and so its image is 
a 3-torsion cocycle of degree equal to at least 8.  A 10-manifold cannot have 3-torsion 
at degree 10, as the degree 10 cohomology is determined entirely by the manifold's 
orientability, which is in $\Z_2$, and so $Q_1$ annihilates 5-classes.  It turns out that $Q_1$ also annihilates 4-classes.  Therefore $Q_1$ may only be nontrivial on a 
3-class $z$.  Physically $z$ may describe the Ramond-Ramond 3-form flux $G_3$ in type IIB or a D6-brane in type IIA which is Poincar\'e dual to $z$.  In the first case $\beta P_3^1z$ is the D-string charge carried by the flux.  In the second the dual of $\beta P_3^1z$ is the singular locus of the D6.  In M-theory, a role for $p=3$ has appeared in \cite{S3}. 



\section{Two Examples}
\label{ex}

\subsection{IIA on a Product of Lens Spaces}

We now recall an example of a nontrivial $Q_1$ action that has appeared in Refs.~\cite{BHK,BRST,Hisham}. Consider the product of lens spaces $X^{10}=S^3/\Z_3 \times S^7/\Z_3$, where the 
$\Z_3$'s are subgroups of the free circle actions on the spheres. T-duality and 
fluxes on similar spaces has been considered in \cite{aussy2005}. We will be interested in the  
cohomology groups of $S^3/\Z_3$ and $S^7/\Z_3$ with $\Z_3$ coefficients. 
These are generated by the 1 and 2-cocycles $x_1$ and $x_2$ for 
$S^3/\Z_3$, and $y_1$ and $y_2$
for  $S^7/\Z_3$. 

The cocycles $x_1$ and $y_1$ do not have integer lifts, as $\beta(x_1)=-x_2$ 
and $\beta(y_1)=-y_2$. However the degree three class
\(
w=x_1y_2-x_2y_1
\)
does admit an integer lift as
\(
\beta(w)=-x_2y_2+x_2y_2=0.
\)

We are interested in the action of $Q_1=-\beta P^1_3$ on $w$, which
gives the non-zero result
\(
d_5(w)=Q_1(w)=-\beta P^1_3(w)=-\beta P^1_3(x_1y_2)=-\beta(x_1y_2^3)=x_2 y_2^3.
\)

Thus the 7-cycle Poincar\'e dual to $w$ is not representable by any nonsingular submanifold, and, as it is not annihilated by the AHSS differential $d_5$, it also does not lift to a K-theory class.  The obstruction is $d_5(w)$ which is dual, inside of the 7-cycle, to a 2-torus.  This suggests that, at least for a certain choice of representatives of the 7-cycle, the singular locus is a 2-torus.  For example a 5-dimensional normal slice to the 2-torus inside of the singular 7-cycle may be the real cone over $\cp^2$.

\subsection{IIA on the Sp(2) Group Manifold}

We have used the AHSS and the critical dimension of type II superstring 
theories to argue that an FW anomaly-free brane lifts to twisted K-theory if and only if the Poincar\'e dual $w$ of the cycle that it wraps is in the kernel of $Q_1=-\beta P^1_3$.  Thom's result on representability is somewhat weaker, while all representable cycles are in the kernel of $Q_1$, there are other obstructions to representability that do not appear in the AHSS.  For example, in Ref.~\cite{BHK} the authors have shown that representability of the dual cycle to $w$ implies 
\(
w\cup P^1_3w=0. \label{rep}
\)

In particular they provided an example in which $\beta P^1_3 w=0$ and 
$w\cup P^1_3 w\neq 0$.  They considered the 10-dimensional group manifold Sp(2).  This is topologically a 3-sphere bundle over a 7-sphere and in particular it has the cohomology ring of the trivial bundle
\(
\H^0(\textup{Sp(2)})=\H^3(\textup{Sp(2)})=\H^7(\textup{Sp(2)})=\H^{10}(\textup{Sp(2)})=\Z.
\)
Notice that the cohomology contains no torsion subgroups and so all of 
the Atiyah-Hirzebruch differentials are trivial and every integral 
cohomology class lifts to a K-theory class.  However the authors prove 
that the generator $w$ of $\H^3=\Z$ does not satisfy the condition 
(\ref{rep}) and so the dual homology class is not representable.  Thus 
in this case there is a K-class which does not correspond to any 
homology class which is 
represented by any nonsingular submanifold.  As every K-class is realizable as a gauge field configuration on some stack of branes, the IIA version of the Sen conjecture implies that such a singular brane configuration must be allowed in IIA string theory.

One may then hope to use the Sp(2) WZW model to provide a nontrivial 
test of the Sen conjecture.  However notice that in the example at hand we have considered a trivial $H$-flux, corresponding to a negative level and therefore a nonunitary conformal field theory.  It may therefore be difficult to decide whether this brane configuration should be allowed.  

\section{Discussion}
\label{dis}

We have argued that an FW anomaly-free brane carries a K-theory charge if it wraps a representable cycle.  We then presented two examples that showed that the nonrepresentability of a FW anomaly-free cycle does not mean that a wrapped brane necessarily does or does not carry a K-theory charge.

In particular we considered one compactification on a product of lens spaces in which a FW anomaly-free brane wrapped on a nonrepresentable cycle does not carry a K-theory charge.  One may object that this is not a legitimate compactification of IIA string theory as it has positive curvature and so is not Ricci flat.  This problem is easily solved.  The product of lens spaces is a $T^2$-bundle over $\cp^1\times\cp^3$ and can be made into
a Ricci-flat space by replacing $\cp^1$ by a Riemann surface of genus greater than 
zero. As this replacement does not affect the calculation of $Q_1$ \cite{BRST}, there will still be a nonrealizable cycle which does not lift to a K-theory class.

One may argue that branes on nonrepresentable cycles are of limited interest, because in string phenomenology one is typically interested in a four-dimensional topologically trivial space times a six, seven or eight-manifold, and all cycles of six, seven and eight-manifolds are representable.  There are two interesting cases in which representability is an issue.  First, one may consider noncritical string theories.  For example nonrepresentable cycles are generic on group manifolds and so in WZW models.  Second, one may consider a spacetime which locally is a product of $\R^4$ or $dS^4$ times a low-dimensional manifold, but in which globally the topology of the four big dimensions is mixed with that of the little dimensions.  For example, at the big bang, at a big crunch and at some horizons the product approximation may fail.  Such a compactification would break four-dimensional Lorentz symmetry a little far away from interesting places like the big bang, and break it a lot at the big bang.  Of course, this is the breaking pattern observed in nature, and for example in the FLRW solution.

It would be interesting to understand the effect of a brane wrapping a nonrepresentable cycle on a low energy effective theory in the remaining dimensions, when there are any.  For example, the $\Z_3$ anomaly that assures that certain nonrepresentable cycles do not yield K-theory charges may correspond to some interesting anomaly in the low energy effective theory.  The low energy physics of D6-branes wrapped on representable cycles that carry torsion K-theory charges has recently been investigated in Ref.~\cite{Marchesano}.

The more interesting question is whether branes can be wrapped on these nonrealizable cycles.  For this we need some description of the worldsheet physics.  We hope that the worldsheet theory on the product of lens spaces is the IR fixed point of a linear sigma model which is just the tensor product of the linear sigma models on the two lens spaces.  This model is somewhat complicated by the strange boundary condition corresponding to the brane which wraps the nonrepresentable cycle.  

In the case of the group manifold, the corresponding WZW model is nonunitary.  It can be made unitary is we add an $H$-flux, but in this would go beyond the scope of our results.  In particular we do not know how $H$-flux changes the AHSS differential $d_5$, although the rational part of the result has been recently provided in Ref.~\cite{AS}.  We hope to use T-duality to find the full expression for $d_5$ in the twisted case.  This would allow one to use a unitary WZW model to determine whether or not a D-brane wrapped on a nonrepresentable cycle can provide a physical boundary condition for fundamental strings.  However, once $H$ corrections are included, it may well be that representability will be replaced with a twisted notion of representability, such as the representability of a section of $PU(\mathcal{H})$ or loop group of $E_8$ bundle over a cycle.


\section* {Acknowledgement}
We would like to thank Peter Bouwknegt, Alan Carey, Varghese Mathai, Tony Pantev, 
Eric Sharpe and Bai-Ling Wang for useful discussions. We also thank the organizers of the 
workshop ``Gerbes, Groupoids, and Quantum Field Theory" at the Erwin Schr\"odinger 
Institute for the invitation and the stimulating environment. 
The work of J.E. is partially supported by IISN - Belgium (convention 4.4505.86), 
by the ``Interuniversity Attraction Poles Programme -- Belgian Science Policy''
and by the European Commission programme MRTN-CT-2004-005104, in which he is
associated to V. U. Brussel. The work of H.S. is supported by an ESI Junior Research 
Fellowship.


\end{document}

1) take beta of w

2) PD of w is $S^1\times S^5/\Z_3$, its the two boundaries that get glued.  It's
also a 2-torus fibered over $\cp^2$.

3) The two cycles of the 2-torus get filled in as in surgery, but the boundaries of
the disks are (3,0) and (0,3)

4) To get the link, move a little bit away from the boundary of the first circle
into the disk, this goes to 3 points in the disk, so you get a $Z_3$ bundle over
something.  The something is the $S^5/\Z_3$, but you don't want the $S^1$ fiber 
because
that is part of the $T^2$ and we're considering an orthogonal slice to the $T^2$.
Quotienting that circle out, depending on which circle you quotient out, you get
either $\cp^2$ of $\cp^2/\Z_3$, I'd guess the second.
Anyway, so you've got a $Z_3$ bundle over one of these, and its continuous.
So in the two cases respectively you'd have to have $\cp^2\times\Z_3$ or
else $\cp^2$, I'd think the second.


It was once believed that, in the absence of D-branes, Ramond-Ramond field strengths were classified by twisted K-theory \cite{MM}.  Evidence for this conjecture came from an analysis of symmetric boundary conditions in the worldvolume theories of open strings \cite{FS}, from the analysis of the Chan-Paton bundles on various unstable D-branes \cite{WittenK} and from the conditions imposed on D-brane embeddings by global worldsheet anomaly cancellation \cite{FW}.  However within a few years, Diaconescu, Moore and Witten realized \cite{DMW} that, in type IIB string theory, the twisted K-theory classification of fluxes is inconsistent with S-duality and so is incorrect.  

With hindsight this was not surprising, as none of the derivations mentioned above is covariant with respect to S-duality.  For example, searching for D-branes as boundary conditions in the conformal field theories will not yield an S-duality covariant classification of branes unless one also includes NS branes and examines the boundary conditions of ($p,q$)-strings, which would be difficult as the fundamental string coupling diverges near an NS5-brane.  Witten's construction of untwisted K-theory from tachyon condensation \cite{Sen} of stacks of D9 and anti D9-branes also could not be S-dualized as D9-branes are already poorly understood, and their S-duals are unknown if they exist at all.  Freed and Witten's worldsheet anomaly was easily S-duality covariantized \cite{me}, however except in the case of 3-form field strengths \cite{DMW,DFM,aussy2005} it was not known how to use it to find a consistency condition for the field strengths.

Some authors have suggested that this inconsistency should be resolved by replacing twisted K-theory with an entirely different generalized cohomology theory \cite{Hisham}.  Instead, in this note we suggest a new approach to this old problem, using the BRST formalism.  In particular, we identify a set of large Ramond-Ramond gauge transformations of the type II supergravity action, essentially those which keep the Wilson loops invariant, and we show that they act nontrivially on the field strengths.  These symmetries act on the integral lattice of de Rham cohomology which satisfies the Dirac quantization condition.  This lattice is often smaller than the full integral cohomology, in which the various field strengths are believed to be valued.  We use the Freed-Witten anomaly to extend the gauge transformations to the full integral cohomology.

We then make a simple observation, which is the central result of this paper.  The BRST cohomology of the RR gauge transformations is isomorphic, as a set, to twisted K-theory.  In particular Atiyah and Hirzebruch have shown that K-theory may be constructed from integral cohomology by taking its cohomology with respect to a series of differential operators.  In other words, a quotient of a subset of cohomology approximates K-theory, and a quotient of a subset of that is a better approximation, and so on.  We argue that Atiyah and Hirzebruch's differential operators are precisely the BRST charges corresponding to the large RR gauge transformations.  In addition the extra, wrong degree, cohomology classes that one adds to the sequence are the ghosts and antighosts.  This is an unusual application of the BRST formalism because the gauge symmetry that we consider is discrete, and so in particular the ghosts and antighosts do not enjoy propagating degrees of freedom.

This all suggests that to find the S-duality covariant classification of RR and NSNS field strengths in IIB, and also the correct classification in IIA, one needs to take the BRST cohomology not only with respect to the RR gauge transformations but also with respect to the NSNS gauge transformations.  This is not easy, as the NS 3-form field strength is itself used in the RR gauge transformations, and as a result these two symmetries cannot be disentangled.  The result appears to be that one does not find that the collection of NSNS and RR field strengths together is a cohomology or even an additive group.  This is no surprise, as the field strengths are solutions to supergravity equations of motion which are nonlinear.  Quotienting by the NSNS transformations one loses more than just the addition however, the S-duality covariant classification in general often has a different cardinality than the twisted K-theory, as was shown in the case of the Klebanov-Strassler geometry in \cite{mecascade}.

We hope that the BRST construction of twisted K-theory will have other applications.  For example, Witten has speculated \cite{strings2000notes} that there may be a twisted K-theory based formulation of type II string theories.  However efforts at constructing such a formulation have been complicated by the fact that twisted K-theory classes are not easily expressed as fields, and so not easily treated using the standard tools of quantum field theories.  However using the current construction one need only start with fields that are ordinary differential forms, supplemented with the Dirac quantization conditions and the various ghost towers that construct a Deligne cohomology and allow the description of torsion classes.  That is, the field content is the same as in $p$-form gauge theory.  Then one finds the gauge symmetries of the path integral, includes ghosts and calculates the BRST cohomology as usual.  

This yields a BRST cohomology which is larger than twisted K-theory.  To obtain twisted K-theory, one must then ignore the values of the various Wilson lines, which are exponentials of the integrals of the gauge connections.  Alternately, one can choose not to ignore these and one would arrive at a differential version of twisted K-theory.  It would be interesting to see if this differential K-theory agrees with that proposed by Freed in \cite{Freed}.

The description of the twisted K-theory of the spacetime or spatial slice $M$ as a BRST cohomology is summarized in the following table.  For concreteness we consider type IIA.
\begin{table}[h]
\begin{center}
\begin{tabular}{|l|l|l|l|}
\hline \textbf{BRST} & \textbf{K-theory} & \textbf{IIA Supergravity} & \textbf{IIA String Theory}\\
\hline fields & $\oplus_k \H^{2k}(M)$ & RR diff. forms $G_{2k}$ & RR int. classes $G_{2k}$\\
\hline ghosts&$\oplus_k \H^{2k+1}(M)$&gauge xforms/branes&gauge xforms/branes\\
\hline BRST operator&$d_{2p+1}$&$d_3=H\wedge$&$d_3=Sq^3+H\cup,\ d_5$\\
\hline Gauge xforms&$d:\HO\rightarrow\HE$&Wilson loop shift&FW monodromy\\
\hline Constraints&$d:\HE\rightarrow\HO$&Bianchi identities&source FW anomaly\\
\hline physical fields&$\K^0_H(M)$&orbits of Bianchi solns&orbits of FW solns\\
\hline anomalies&$\K^1_H(M)$&stable p-branes&stable D-branes\\
\hline
\end{tabular}
\end{center}
\caption{K-theory vs BRST in type IIA supergravity and in type IIA string theory}
\end{table}
Stable, consistent D-branes correspond to anomalies, despite the fact that their partition functions are gauge invariant.  Indeed they appear in the supergravity equations of motion as violations of the conservation of RR current.  They also correspond to anomalies in the sense of Ref.~\cite{EF} as instantonic D-branes which form, sweep nontrivial cycles and then decay change the cohomology class of the fluxes that they source, and therefore mediate a tunneling between different source-free RR solutions \cite{MMS}.  Notice in particular that on their worldvolumes the unimproved field strengths are not defined and so the gauge transformations are not defined, similarly to the worldvolume of a magnetic monopole in QED where the gauge potential $A$ is not defined.  In the quantum theory this may correspond to a nontrivial inner product between states with stable D-branes and states with Ramond-Ramond fields that are pure gauge.  The fact that D-branes correspond to elements of the BRST cohomology and in particular are BRST closed implies a sort of Wess-Zumino consistency condition which is enforced classically by the supergravity equations of motion and quantum mechanically by the Freed-Witten (FW) anomaly.

In section~\ref{classsec} we will use the democratic formulation \cite{T,Townsend,VanProeyen} of classical supergravity to derive the large RR gauge transformations of the components of the RR field strengths that are valued in an integral sublattice of the de Rham cohomology.  In section~\ref{ahsssec} we will describe Atiyah and Hirzebruch's construction of K-theory from cohomology, and show that restricted to a lattice of de Rham cohomology this yields the gauge-invariant field strengths found in the supergravity analysis.  Finally, in section~\ref{fwsec} we will use Maldacena, Moore and Seiberg's interpretation \cite{MMS} of the Freed-Witten anomaly to extend the action of the large RR gauge transformations to the full integral cohomology, that is, we will find the action on torsion-valued fields.  We will argue that the corresponding BRST operator is the Atiyah-Hirzebruch differential on both nontorsion and $\Z_2$ torsion cohomology classes.  In type IIB the Freed-Witten anomaly needs to be generalized to capture the $\Z_3$ torsion contributions.  An example of a brane with a $\Z_3$ generalized Freed-Witten anomaly is provided.

\section{Large gauge transformations} \label{classsec}

In this section we will review the large gauge transformations of the gauge potential in QED, these are gauge transformations that shift the integrals of the gauge potential by integers and so leave the Wilson loops invariant.  We will then find the corresponding large RR gauge transformations in the democratic formulation of type II supergravity theories and argue that, unlike the case of QED, these transformations also change the field strengths.  We present a classification of gauge equivalence classes of those field strengths that satisfy the RR supergravity equations of motion and Dirac's quantization condition.

\subsection{A Warm up: QED}

In the absence of charged matter, Maxwell's theory of electromagnetism in any number of dimensions is described by the Lagrange density
\beq
{\mathcal L}=F\wedge\star F
\eeq
where $F$ is a 2-form field strength containing the electric and magnetic field components.  As there is no matter, this field strength is the only observable.  To define the classical theory we need to supplement this Lagrangian with an additional equation, for example we may impose that the field strength is closed
\beq
dF=0. \label{bianchi}
\eeq
In this case on any local patch of spacetime we may define a one-form gauge potential $A$ by demanding that
\beq
F=dA. \label{a}
\eeq
Then Eq.~(\ref{bianchi}) becomes a Bianchi identity for $A$, which is the condition that $A$ is locally well-defined and so is annihilated by the square of the exterior derivative $d$.

Eq.~(\ref{a}) does not uniquely specify the gauge potential $A$, but only specifies it up to an additive shift $\eta$ which is called a gauge transformation.  One often says that $\eta$ may be any exact one-form.  However, if the spacetime topology is nontrivial then in the absence of charged matter $\eta$ is not necessarily exact but only needs to be closed: $d\eta=0$.  For example, if the spacetime is a circle parameterized by the coordinate $\theta$ then one may shift $A$ by $\eta=cd\theta$
\beq
A\mapsto A+\eta=A+cd\theta \label{const}
\eeq
where $c$ is any nonzero constant.  $\eta$ is then closed but not exact.  The transformation (\ref{const}) leaves invariant the field strength $F$ and therefore also the Lagrange density $\mathcal L$, identifying (\ref{const}) as a gauge transformation.

If one includes electrically charged matter in the theory then the set of allowed gauge transformations is restricted.  For example, one may couple the theory to a conserved current $\mathcal J$ of particles with charge $q$ by introducing the following Lagrange density
\beq
{\mathcal L}=F\wedge\star F+q\mathcal JA. \label{l2}
\eeq
Again one may consider gauge transformations of the form $A\mapsto A+\eta$ and try to find the values of $\eta$ for which the Lagrange density is invariant.  The current $\mathcal J$ is an arbitrary closed 3-form and so it is necessary that the gauge transformations separately preserve both terms in the Lagrangian (\ref{l2}).  We have already found that $\eta$ needs to be closed to preserve the first term.  The second term depends explicitly on $A$ and so will not be invariant.  

We are searching for a condition on $\eta$ such that both the field strength $F=dA$ and the action $S=\int_M \mathcal L$ are gauge-invariant.  In particular, for an orientable spacetime $M$ one may use the fact that $\mathcal J$ is closed for any conserved current to perform the integral over all spacetime directions except for one, leaving $S=q\int_N A$, where $N\subset M$ is the Poincar\'e dual of $\mathcal J$, which is the worldline of an electrically charged particle.  The action will therefore be invariant precisely when $\eta$ is exact.

We have imposed too strong of a condition on $\eta$.  The action of a quantum theory is not an observable, and does not need to be well-defined.  Instead only the path-integral measure 
\beq
e^S=e^{\int_M F\wedge\star F}e^{q\int_NA} \label{part}
\eeq
must be well-defined, where the first factor on the right hand side is well defined when $\eta$ is closed.  This means that those one-forms $\eta$ that shift $q\int_NA$ by an integer are also gauge transformations.  In other words $\eta$ does not need to be exact, but may represent any cohomology class that lies on an integer sublattice $\Z^k$ of the first real cohomology of the spacetime.  For example, if the spacetime is a 2-torus $T^2$ then the allowed gauge transformations, up to an exact one-form, are
\beq
\eta=md\theta+nd\phi\in\Z^2\subset\textup{H}^1(T^2)=\R^2
\eeq
where $m$ and $n$ are integers and $\theta$ and $\phi$ are the coordinates of two cycles that span the torus.  $\Z^2$ is then said to be the group of large gauge transformations.  These gauge transformations are called ``large'' because only the identity itself is in the same connected component of the gauge group as the identity.

These large gauge transformations are used, for example, in the construction of the Dirac string.  The gauge potential $A$ is not globally defined on a 2-sphere that links a magnetic monopole, however it may be expressed on two patches, one of which intuitively contains the intersection of the Dirac string with the sphere.  On the overlap of the two patches the value of $A$ is defined on both patches, but the values on the two patches do not agree.  Instead the values on the overlap are related by a gauge transformation which is a large gauge transformation.  In this case the large gauge transformation is a transition function which is only defined on the overlap, therefore it is classified by the first cohomology of the overlap, which is related to the second cohomology of the whole spacetime via the Mayer-Vietoris sequence.  It is the second cohomology of the spacetime that classifies nontrivial gauge bundles, such as the bundle on the sphere linking a monopole.

An alternate interpretation of the large gauge transformations is as follows.  The second factor in the action (\ref{part}) is a Wilson loop, normalized by the electric charge of the fundamental particles.  In general, transformations of $A$ by closed forms $\eta$ change the Wilson loops, which leads to a measurable Berry's phase when two charged identical particles encircle $N$ in different directions and interfere.  Gauge transformations, by definition, leave the observables invariant.  Thus the Berry's phase must be invariant, and so the Wilson loop must be invariant.  This is the case precisely for gauge transformations on the above integral lattice.

\subsection{Torsion gauge transformations}

This subsection is more difficult to read than the previous and one may skip it and still understand the main idea of sections \ref{classsec} and \ref{ahsssec}.

The main idea in this subsection, and in section \ref{fwsec}, is that the field content of string theory lies not only in the values of the supergravity fields, but also in choices of phases in the path integral, which are invisible in supergravity.  While closed differential forms that satisfy Dirac's quantization condition are classified by an integer lattice of de Rham cohomology, both fields and phases are classified by integral cohomology.  An integral cohomology class $\H^p(M)$ of the space $M$ may always be written as the direct sum of two terms.  The first term is called the free part, and is an integer lattice $\Z^k$ where $k$ is called the $p$th Betti number and is equal to the rank of the $p$th de Rham cohomology group
\beq
\H_{\rm{de\ Rham}}^p(M)=\R^k\hsp \H^p(M)=\Z^k\oplus_i \Z_{q_i}.
\eeq
The torsion part is the direct sum of cyclic groups $\Z_{q_i}$, which are the additive groups of integers modulo $q_i$.  The supergravity field data is then classified by the free part of the integral cohomology, which is isomorphic to an integer lattice of the de Rham cohomology.  However in the full quantum theory we must also consider the torsion part of the cohomology, corresponding to the aforementioned phases.  The goal of sections \ref{fwsec} will be to extend the action of our BRST operator to the torsion part of the integral cohomology.

Including torsion terms, there are more gauge transformations than just the integral lattice described above.   It is believed that there is a configuration of the gauge field corresponding to every topological circle bundle with connection $A$.  The topologies of circle bundles are in one to one correspondence with classes of the second integral cohomology via an isomorphism called the first Chern class.  In particular, if $A$ is considered, as usual, to be a one-form on each coordinate patch, then the transition functions of $A$ do not necessarily uniquely identify the topology of the gauge bundle.  Physically this corresponds to the fact that the sector of a gauge theory is not determined entirely by the transition functions of $A$, but also by the transition functions of the charged matter. 

For example, consider a spacetime which is topologically the projective space $\rp^2$, which is the quotient of the two-sphere $S^2$ by the $\Z_2$ antipodal map.  This spacetime needs to be Euclidean, but the signature is not important for these considerations and one could think of the $\rp^2$ as a timeslice in a Minkowski spacetime.  The second integral cohomology group of $\rp^2$ is the cyclic group of order two
\beq
\H^2(\rp^2)=\Z_2 \label{z2}
\eeq
and so we expect this theory to contain two topological sectors corresponding to the gauge bundles whose Chern classes are the two elements in (\ref{z2}).  

$\rp^2$ can be visualized as the northern hemisphere of the two-sphere $S^2$ with a 2-fold antipodal identification of the equator.  Then as one descends southward to the equator in the eastern hemisphere, one finds oneself traveling northwards from the equator in the western hemisphere.  A QED configuration on this spacetime consists of a circle bundle over the $\rp^2$, and the charged matter fields are sections of an associated bundle.  These bundles can be trivialized over the northern hemisphere, thus all of the topological information is contained in the transition function on the equatorial circle, that is, one needs to know what happens to the field as one passes through the equator and jumps to the other side of the sphere.  The first de Rham cohomology group is trivial, therefore all closed one-forms $\eta$ are also exact and so correspond to transitions of $A$ that may be continuously deformed away.  Thus the information about the topological sector of the theory is not captured by a one-form $\eta$ as in the previous subsection.

Not all transformations of the matter field may be continuously deformed away.  In particular, one may consider a gauge transformation in which the matter field changes sign as one crosses the equator, which is similar to the sign change of a fermion when one rotates once all of the way around the SO(3)$\cong\rp^3$ subgroup of the Lorentz group.  No more general phase rotation is possible, as the fundamental group of $\rp^2$ is $\Z_2$ and so a loop that crosses the equator twice may be deformed away and so cannot have a topological phase rotation, thus the square of the transition function is the identity.  This transformation is nontrivial, and corresponds to the only nontrivial circle bundle on $\rp^2$, that is, to the bundle with Chern class equal to the element $1\in\Z_2$ in (\ref{z2}) and not the identity element $0\in\Z_2$.  

The $\rp^2$ example illustrates that in general there are physically inequivalent field configurations which do not correspond to any transition functions of $A$.  In fact, the field configurations always correspond to the second integral cohomology.   The free part can described by locally defining the gauge potential $A$ and then attaching these patches using transition functions $\eta$ in the integral lattice described above.  The torsion terms $\Z_k$, on the other hand, are defined to be choices of phase in the path integral depending on the topology of the nilpotent world line of the electric particle.  A nilpotent world line $c$ is one that corresponds to a torsion term $\Z_k$ in the first integral homology group,  which means that the trajectory $kc$, which winds around the loop $c$ $k$ times, is a boundary and so may be deformed away.  For example, if one encircles the noncontractible circle in $\rp^3$ twice, then one traces out a contractible path.  

The different possible phases in the path integral need to be dual to these $\Z_k$'s, and the universal coefficient theorem implies that the torsion part of the second cohomology is equal to the torsion part of the first homology and so classifies these phases.  In particular, if the trajectory encircles a $k$-nilpotent cycle $j\in\Z_k$ times, and one is in the $i$th topological sector of the gauge theory, then the phase is given by $ij$.

Each choice of the path integral phases corresponds to a bundle which in turn corresponds to a choice of transition function of the gauge or matter fields on the overlaps of pairs of patches.  Thus each bundle corresponds to a gauge transformation and the classification of bundles by first Chern classes is also a classification of gauge transformations.  Chern classes, as in the $\rp^2$ case, are not necessarily valued in the integral sublattice of the real cohomology, which is isomorphic to the free part of the integral cohomology.  But rather, Chern classes are arbitrary elements of the integral cohomology.  This suggests that the group of possible gauge transformations is not only the free part of the integral cohomology described in the previous subsection, but apparently is the full integral cohomology.  Of course we do not claim that this observation is new, it goes back at least 20 years \cite{Alvarez}.

Thus the gauge field configurations are classified by the first Chern class which is an arbitrary element of the second integral cohomology whose the free part is captured by a two-form called the field strength.  In what follows, we will often refer to the entire integral class as the field strength despite the fact that the torsion information is encoded in the transition functions of the matter fields and not in the 2-form $F$ itself.

We will be interested in gauge transformations that are defined on the whole spacetime, and not just on overlaps.  Such gauge transformations correspond geometrically to changes of trivializations of bundles.  It is these transformations that relate physically equivalent configurations.  The above argument suggests that these are classified by the classes in the full first integral cohomology.  Due to the universal coefficient theorem, this is isomorphic to the integral lattice of de Rham cohomology that we considered in the previous subsection.  However this isomorphism fails at the higher dimensions that will be relevant to our discussion of supergravity below.  While these gauge transformations can change the gauge connection and some phases in the definition of the partition function, they do not affect the field strength in QED.  On the other hand, in IIA supergravity we will see that the field strengths are no longer invariant under the large gauge transformations.  We will argue that the gauge-invariant quantities are not differential forms, but rather they are classes in twisted K-theory.

\subsection{Type II supergravity}

For concreteness we will restrict our attention to type IIA supergravity, but to obtain type IIB one need only shift the dimensions of the fields by one in either direction.  Massless type IIA supergravity contains two observable Ramond-Ramond $p$-form field strengths, a 2-form $F_2$ and a 4-form $F_4$.  The Lagrangian density contains the usual kinetic terms for these fields
\beq
{\mathcal L}\supset F_2\wedge\star F_2+F_4\wedge\star F_4.
\eeq
We will use the democratic formulation of type II supergravity \cite{T,Townsend,VanProeyen} in which there are no Chern-Simons terms, and it is conventional to give new names to  the Hodge dual field strengths
\beq
F_6=*F_4\hsp F_8=*F_2.
\eeq
Furthermore we will define a single form $F$, called the improved field strength, which is the formal sum of all of the field strengths.  Notice that $F$ does not have a definite degree.

As in the case of QED, the Lagrangian density does not suffice to determine the theory.  One must also impose a constraint on the field strengths.  In the case of supergravity the constraint is a nontrivial generalization of the Bianchi identity in QED
\beq
(d+H)F=0 \label{sbianchi}
\eeq
where $H$ is a closed 3-form known as the NSNS field strength.  $(d+H)$ squares to zero as $dH=0$ and so the $(d+H)$-closure of $F$ in the generalized Bianchi identity (\ref{sbianchi}) implies that locally $F$ is $(d+H)$-exact.  This means that on local patches of spacetime we may define a potential $C$, which is a form of mixed odd degrees such that
\beq
F=(d+H)C. \label{sa}
\eeq
Then Eq.~(\ref{sbianchi}) becomes a Bianchi identity for $C$, which is the condition that $C$ is locally well-defined and so is annihilated by the square of the nilpotent operator $d+H$.  We will often decompose $C$ into its component $p$-forms $C_1$, $C_3$, $C_5$ and $C_7$.

Eq.~(\ref{sa}) does not uniquely specify the gauge potentials $C$, but only specifies them up to an additive shift
\beq
C\mapsto C+\eta
\eeq
which is called a gauge transformation.   As in QED if the spacetime topology is nontrivial then in the absence of charged matter $\eta$ is not necessarily $(d+H)$-exact but only needs to be $(d+H)$-closed: $(d+H)\eta=0$.  

There is a second natural definition of field strength in this theory
\beq
G_{p+1}=dC_p.
\eeq
As $C$ is locally well-defined, this field strength will satisfy the usual Bianchi identity $dG=0$ and so its integral will be used to measure D-brane charge, therefore unlike $F$ it will be quantized.  However $G$ is not gauge-invariant, instead 
\beq
G_{p+1}\mapsto G_{p+1}+d\eta_{p}= G_{p+1}-H\wedge \eta_{p-2} \label{sgauge}
\eeq
where in the second equality we have used the $(d+H)$ closure of $\eta$.
We may use the Bianchi identity to find a relation between $G$ and $F$
\beq
0=dG=dF+H\wedge G.
\eeq
As $F$ is gauge-invariant it is not transformed under transition functions and so it is globally defined.  Thus $dF$ is exact and so $H\wedge G$ is a trivial element of de Rham cohomology.  The triviality of $H\wedge G$ in cohomology will play the role of the constraints in the BRST interpretation.

Analogously to the case of QED, type II supergravity without matter is invariant under gauge transformations with gauge parameter equal to any $(d+H)$-closed $\eta$.  However if we add charged matter to the theory than we must check that the resulting phase shift in the path integral is well-defined, and this will impose a further restriction on $\eta$.  Charged matter in this case consists of $p$-branes, which in classical supergravity can wrap any cycle $N\subset M$ such that
\beq
\int_N H=0. \label{noh}
\eeq
The contribution of such a wrapping to the Wess-Zumino terms of the $p$-brane worldvolume path integral is
\beq
S_{WZ}=e^{\int_N e^{F+B}C}
\eeq
where the closed 2-form $F$ is the field strength of the worldvolume $U(1)$ gauge field and $B$ is the pullback of the NS 2-form, which is well-defined on $N$ because of Eq.~(\ref{noh}).  Under the gauge transformation $C\mapsto C+\eta$ the Wess-Zumino term transforms as
\beq
S_{WZ}=e^{\int_N e^{F+B}C}\mapsto e^{\int_N e^{F+B}(C+\eta)}=e^{\int_N e^{F+B}\eta}S_{WZ}. \label{sxform}
\eeq
Thus gauge invariance imposes that
\beq
\int_N e^{F+B}\eta\in\Z \label{nvinc}
\eeq
for every $N$ satisfying (\ref{noh}).  

At first (\ref{nvinc}) may seem impossible to satisfy, as $\eta$ is not closed under $d$ but only under $d+H$ and so it may appear as though the integral depends on the choice of homological representative $N$.  However this is not the case, since the integrand is closed
\beq
d(e^{F+B}\eta)=(de^F)e^{B}\eta+e^F{de^{B}}\eta+e^{F+B}d\eta=(dF)e^{B}\eta+e^{F}He^B\eta-e^{F+B}H\eta=0
\eeq
where we have used the closure of $F$ and the fact that $H$ commutes with even forms.  Thus on each cycle $N$ the integrand in the phase (\ref{sxform}) is closed.  As each point in $M$ is in some such cycle, the integrand is everywhere closed and so in particular can be lifted, although not canonically, to a cohomology class on $M$.  

The quantization condition (\ref{nvinc}) is not sufficient to impose that the resulting cohomology class is integral, as it applies only to cycles $N$ that do not support $H$ flux.  However, if one projects out the image of the operator $(H\wedge)$ from the de Rham cohomology group then the image of the lift of the gauge transformation under this projection will inhabit an integral lattice of the resulting quotient of the de Rham cohomology.  In fact, to compute the gauge transformations of $G$ in (\ref{sgauge}) we are not interested in gauge transformations $\eta$ in the image of $(H\wedge)$, because $G$ shifts by $H\wedge\eta$ which is invariant under shifts $\eta\mapsto\eta+H\wedge\Lambda$.  Thus the classes $\eta$ in the gauge transformation of $G$ may be taken to be integral classes.  In fact, the lift of the phase in (\ref{sxform}) from the integral cohomology of $N$ to that of $M$ is not even well-defined on image of $(H\wedge)$, as one needs to paste together the various values of the $B$-field with NS gauge transformations, and so we are fortunate that in considering the gauge transformations of $G$ we do not need to know the component of $\eta$ that is in the image of $H\wedge$.  If one wishes to extend this note to find the gauge orbits of $C$ then one will again be confronted with this problem.


In conclusion, it appears that, as in the case of QED, in type II supergravities there are large gauge transformations in which the gauge parameter $\eta_p$ lie, up to a correction in the image of $(H\wedge)$ which does not contribute to the transformation of $G$, on an integral lattice of the $p$th de Rham cohomology group.  Similarly, in the quantum theory there will also be torsion terms that can be included by allowing $\eta_p$ to be an arbitrary element of the $p$th integral cohomology group.  Ignoring the torsion for now, one may compute the groups of allowed, gauge-inequivalent field strengths $G_{p+1}$.  We are not going to be interested in the connections $C_p$, although including them would be interesting and would hopefully lead to a kind of differential K-theory.  

We have seen that the equations of motion imply that $H\wedge G_{p+1}$ represents the trivial cohomology class, thus the allowed field strengths form the kernel of the operator
\beq
H\wedge:\textup{H}^{p+1}(M)\rightarrow\textup{H}^{p+4}(M):x\mapsto H\wedge x. \label{solutions}
\eeq
The gauge transformations $\eta_p$ are classified by the $p$th cohomology.  In fact we'll be interested in $\eta_{p-2}$, which is classified by the $(p-2)$nd cohomology.  These act on the field strengths via Eq.~(\ref{sgauge}), which adds $H\wedge \eta_{p-2}$ to $G_{p+1}$.  As $\eta_{p-2}$ may be any ($p-2$)nd cohomology class, $G_{p+1}$ is only defined up to $(p+1)$-classes which are the wedge product of $H$ with something.  That is, the possible shifts of $G_{p+1}$ are the image of the operator
\beq
H\wedge:\textup{H}^{p-2}(M)\rightarrow\textup{H}^{p+1}(M):x\mapsto H\wedge x. \label{xforms}
\eeq
Quotienting the solutions of the RR supergravity Bianchi identities (\ref{solutions}) by the RR gauge transformations (\ref{xforms}) gives a classification of gauge-inequivalent RR field strengths
\beq
G_{p+1}\in \frac{\textup{Ker}(H\wedge:\textup{H}^{p+1}(M)\rightarrow\textup{H}^{p+4}(M))}{\textup{Im}(H\wedge:\textup{H}^{p-2}(M)\rightarrow\textup{H}^{p+1}(M))}. \label{sugraclass}
\eeq

The classification of field strengths (\ref{sugraclass}) has an interpretation in terms of the BRST cohomology with respect to the large gauge transformations (\ref{xforms}) where $\eta$ is valued in integral cohomology.  The numerator consists of the field strengths that satisfy the constraints, which are given by the Bianchi identities above, where as the denominator consists of those field strengths that are pure gauge.  On the other hand, no physical fields occupy the cohomology classes $\textup{H}^{p-2}$ and $\textup{H}^{p+4}$, which are even and so have the wrong statistics to be field strengths in type IIB.  Yet the introduction of these two classes is crucial, $\textup{H}^{p-2}$ because it is the preimage of the pure gauge field strengths under the $H\wedge$ map (\ref{xforms}) and $\textup{H}^{p+4}$ because it is the image of the forbidden field strengths under the $H\wedge$ map (\ref{solutions}).  This leads us to identify $\textup{H}^{p-2}$ and $\textup{H}^{p+4}$ as topological (anti)ghost fields and $H\wedge$ as the BRST operator.

Notice that we have not imposed the NSNS equations of motion, nor have we quotiented by the field strengths that may be removed by NSNS gauge transformations.  Thus we have only partially classified the field strengths in IIB supergravity.

\section{The Atiyah-Hirzebruch spectral sequence} \label{ahsssec}

\subsection{Adding torsion}

In the last section we used supergravity to classify gauge orbits of Ramond-Ramond field strengths that satisfy the Bianchi identities.  We argued that the field strengths should live in integral cohomology, which is the sum of a free part of the form $\Z^j$ plus a torsion part that consists of cyclic groups $\Z_{q_i}$.  The free part is equal to the integral lattice of the de Rham cohomology.  Our supergravity techniques are rather limited, in that we use a classical action and try to obtain information about a quantum theory.  In particular, in the classical limit the torsion part of the cohomology vanishes and so we cannot simply use the gauge symmetries of the supergravity action to find the action of the BRST operator on the torsion terms.  This means that the classification (\ref{sugraclass}) of RR field strengths will not be quite right on a spacetime whose integral cohomology contains a nontrivial torsion piece.  In section \ref{fwsec} we will use the Freed-Witten anomaly to recover these missing torsion contributions.

In the present section we will find the torsion contributions from an entirely different perspective.  We will review the Atiyah-Hirzebruch spectral sequence (AHSS) construction of twisted K-theory, and show that if the torsion pieces of the cohomology are simply dropped then instead of constructing twisted K-theory, one arrives at the above classification of RR field strengths (\ref{sugraclass}).  In other words, we will prove that our above classification of supergravity configurations is an approximation of twisted K-theory in which one ignores the torsion parts of the integral cohomology, taking only the free part which is a sublattice of de Rham cohomology.  In particular, if we consider a theory with a spacetime whose cohomology contains no torsion, such as the SU(2) WZW model, then the classification (\ref{sugraclass}) is already correct as we will see in Subsec.~\ref{wzwsec}.  This, in turn, will provide us with a conjecture for the torsion corrections to the supergravity results of the previous section.  One need only reintroduce the torsion terms that appear in the AHSS.  In section \ref{fwsec} we will argue that those torsion terms are in fact required for the cancellation of global anomalies on the worldsheets of open fundamental strings.

\subsection{What kind of objects are twisted K-groups?}

The integral cohomology $\textup{H}^*(M)$ of a space $M$ is a collection of abelian groups $\textup{H}^k(M)$ indexed by a nonnegative integer $k$ and endowed with a product
\beq
\cup:\textup{H}^j(M)\otimes_\Z \textup{H}^k(M)\rightarrow \textup{H}^{j+k}(M):x\otimes y\mapsto x\cup y \label{mult}
\eeq
called the cup product.  The free part $\Z^k$ of the integral cohomology group $\Z^k\oplus_i\Z_{q_i}$ is isomorphic to an integral lattice of de Rham cohomology $\R^k$.  On this integral lattice the cup product $x\cup y$ reduces to the wedge product of differential forms $x\wedge y$.  Thus we may guess that the wedge products of the previous section will, when we include torsion, be written as cup products plus torsion corrections.

The twisted K-theory $\K_H^*(M)$, with twist $H$, of a space $M$ is another collection of abelian groups $\K_H^k(M)$, this time indexed by a general integer $k$ and depending on an integral three-class $H$.  However, unlike the cohomology groups, the $\K$ groups are not independent.  Instead they are related by the Bott periodicity relation
\beq
\K_H^j(M)\cong \K_H^{j+2}(M)
\eeq
and thus it will suffice to compute $\K_H^0(M)$ and $\K_H^1(M)$.  This structure better mimics the charge structure of D-branes.  For example there are dielectric and fractional D$p$-branes that carry a half unit of D$(p-2)$-brane charge, meaning that twice the generator of D$p$ charge should be equal to the generator of D$(p-2)$ charge.  Such a relation, between generators of different degrees, would be impossible in cohomology.  However in K-theory it is automatic, a single group classifies all even degrees, while another classifies all odd degrees.  Thus, in each string theory, one K-group classifies all of the D-branes simultaneously.  As D-branes source fluxes, one can arrive at a similar story for fluxes.

While ordinary K-theory admits a multiplication similar to (\ref{mult}), the K group twisted by a fixed $H$ admits no such multiplication.  Instead when one multiplies two twisted K-groups the twists add.  However the twist corresponds to a fixed NS 3-form flux, and so this multiplication changes the $H$ flux and so does not correspond to any obvious physical process in string theory.  Setting the twist of one of the factors to zero, one finds that the twisted K-theory $\K^*_H(M)$ is a module of the untwisted K group $\K^0_H$, but no physical interpretation of this fact in string theory has appeared to date.

\subsection{Constructing twisted K-theory}

We will now describe the AHSS, which is an algorithm for computing the twisted K groups $\K^*_H(M)$ from the $H$ flux and the integral cohomology $\H^*(M)$.  In type IIB string theory RR field strengths correspond to odd cohomology classes, and so we shall see they will be classified by $\K^1_H(M)$.  Similarly in IIA string theory the fluxes are even classes and so will be described by $\K^0_H(M)$.

We first assemble the even and odd cohomologies into two big groups $\HE$ and $\HO$
\beq
\HE(M)=\oplus_k \H^{2k}(M)\hsp \HO(M)=\oplus_k \H^{2k+1}(M).
\eeq
We want to construct the twisted K-group $\K^i_H(M)$ by finding a finite set of improving approximations $E^i_j$ starting with the odd cohomology and finishing with a set $E^i_n$ of the same cardinality as $\K^i_H(M)$
\beq
E^0_1=\HE(M)\hsp E^1_1=\HO(M)\hsp |E^i_n|=|\K^i_H(M)|.
\eeq
In general $E^i_n$ will not be isomorphic to $\K^i_H(M)$ as a group, that is, the addition rule will be different.  The addition rules are related by an extension problem.  However, as there is no addition rule for fluxes in the S-duality covariant case anyway, since the sum of two sets of fields that satisfy the nonlinear supergravity equations of motion generically is not another solution, we will not concern ourselves with reproducing the addition rule in this note.

To get from $E^i_1$ to $E^i_n$ we will need to introduce a series of differential operators $d_{2j+1}$ which are degree $(2j+1)$ cohomology operations, in other words
\beq
d_{2j+1}:\H^k(M)\rightarrow\H^{k+2j+1}(M)\hsp d_{2j+1}d_{2j+1}=0.
\eeq
As we will soon see, only the first differential operator, $d_3$, needs to be well-defined on the full integral cohomology.  To pass from $E^i_j$ to the next approximation $E^i_{j+1}$ we need to take the cohomology with respect to the differential operator $d_{2j+1}$
\beq
E^i_{j+1}=\frac{\Ker(d_{2j+1}:E^i_j\rightarrow E^{i+1}_j)}{\Im(d_{2j+1}:E^{i+1}_j\rightarrow E^{i}_j)}.    
\eeq
In particular, we see that while the differentials do not need to be defined on all of the cohomology classes, as some classes will be eliminated earlier in the procedure by the kernel operations, they do need to be well-defined on the equivalence classes obtained by quotienting by the images of their predecessors.  This identifies the differentials beyond $d_3$ not as ordinary cohomology operations, but as secondary cohomology operations obtained perhaps from Toda brackets of primary cohomology operations.  However the details of the constructions of these $d$'s will not concern us, it will suffice to use the claims of Diaconescu, Moore and Witten \cite{DMW} and then Maldacena, Moore and Seiberg \cite{MMS} that they compute Freed-Witten anomalies.

In known examples, the first differential $d_3$ is the only differential that does not vanish in the absence of torsion cohomology.  In fact, the torsion free parts of all of the differentials have been computed in \cite{AS}, where they have been seen to be Massey products of $H$.  This implies in particular that they will always be trivial on compact Kahler manifolds.  However in principle they could appear in supergravity, and it would be interesting to find the corresponding gauge transformations.  $d_3$ may be written
\beq
d_3x=Sq^3x+H\cup x
\eeq
where $H$ is our familiar NS 3-class and $Sq^3$ is a cohomology operation known as a Steenrod square which takes an integral class in the $k$th cohomology to a class in the $(k+3)$rd cohomology, as does the cup product with $H$.  Unlike the cup product with $H$, however, $Sq^3$ is only nontrivial when acting on $\Z_2$ torsion components of $\H^k(M)$, and the image is likewise always a $\Z_2$ torsion component of $\H^{k+3}(M)$.  This means in particular that in the supergravity limit such torsion components disappear because there is no Dirac quantization and so $\Z_2$ torsion fields may always be written as two times another field and so torsion fields are even and therefore equal to zero modulo 2.  Therefore in the classical supergravity limit the $Sq^3$ term vanishes and $d_3$ reduces to the operator in Eq.~(\ref{solutions}).  

The physical interpretation is that the $Sq^3$ term measures an obstruction to a brane worldvolume being $spin^c$.  If the worldvolume is not $spin^c$, then the fermion partition function will be anomalous unless this anomaly is canceled by the $H$ flux.  Geometrically this cancellation may be realized by tensoring a $spin^c$ bundle with 3-class $W_3$ by an $LE_8$ bundle with a three-class that cancels $W_3$, thus the tensor product bundle has no 3-class and so no obstruction to choosing a $spin$ structure.

This anomaly, which comes from an ill-definedness in the square root of a determinant of a Dirac operator in the path-integral measure, is not visible in the classical theory.  Thus in the classical limit all that remains of the above procedure is $d_3$ which is equal to the wedge product with $H$.  One then recovers the classification (\ref{sugraclass}).

The appearance of (\ref{sugraclass}) as the classical limit of the AHSS construction suggests the conjecture that the AHSS construction provides a quantum completion of the supergravity construction in the previous section.  In particular, one may conjecture that the quantum-corrected Bianchi identities imply that physical states are in the kernel of the AHSS differentials $d_{2j+1}$.  In addition, one may conjecture that the quantum-corrected gauge transformations are such that the images of the AHSS differentials are pure gauge.  Now the $K^{i+1}_H$'s are interpreted as a series of nonpropagating ghosts, and the $d_{2j+1}$'s as a series of BRST operators.  Of course, in string theory only $d_3$, $d_5$ and perhaps $d_7$ may ever be relevant for dimensional reasons.

\subsection{An example: The twisted K-theory of $S^3$} \label{wzwsec}

The supersymmetric SU(2) WZW model at affine level $k-2$ describes string theory on the three-sphere $S^3$ with $k\neq 0$ units of NS 3-form flux
\beq
\int_{S^3} H=k.
\eeq
The integer cohomology of the three-sphere is
\beq
\H^0(S^3)=\H^3(S^3)=\Z\hsp \H^1(S^3)=\H^2(S^3)=0
\eeq
where $0$ is the trivial group, which contains only the identity element.  In particular, the cohomology contains no torsion and so $Sq^3$ acts trivially.  As the maximum difference in the dimensions of two elements of the cohomology is equal to three, the operators $d_{2j+1}$ are trivial for $j>1$, leaving only $d_3$, which contains only the cup product with $H$.  

If $e_0$ is the generator of $\H^0(S^3)=\Z$ and $e_3$ is the generator of $\H^3(S^3)=\Z$ then
\beq
d_3 e_0=H\cup e_0 = k e_3\hsp d_3 e_3=H\cup e_3=0
\eeq
where $H$ kills $e_3$ because the cup product of two 3-classes is a 6-class, but the 6-cohomology is trivial.  Thus $d_3$ acts nontrivially on all of $\H^0(S^3)$, but annihilates all of $\H^3(S^3)$.  In other words, the kernel of $d_3$ is the third cohomology group, $\Z$.  Similarly the image of $d_3$ consists of those elements $k\Z\subset\Z$ of $\H^3$ which are multiples of $k$.  Note that both the kernel and the image lie in $\H^3$, which is in $\HO$ because $3$ is odd.  $\HE$, on the other hand, contains no elements of the kernel and so $\K^0_H(S^3)$ is trivial.  Summarizing, we have found that
\beq
\K^1_H(S^3)=\frac{\Ker(d_3:\HO(M)=\Z\rightarrow \HE(M)=\Z)}{\Im(d_{3}:\HE=\Z\rightarrow \HO=\Z)}=\frac{\H^3(M)=\Z}{k\H^3(M)=k\Z}=\Z_k. \label{wzw}
\eeq
As was shown in Ref.~\cite{FS}, the twisted K-groups (\ref{wzw}) reproduce the known symmetric D-branes in the supersymmetric SU(2) WZW model.
    
Physically the nontrivial element $j\in\Z_k=K^1_H(S^3)$ corresponds to a RR 3-form field strength $G_3$ such that $\int_{S^3}G_3=j$ in an embedding of the SU(2) WZW model into IIB string theory, for example on a 3-sphere linking an NS5-brane.  We recall that $G_3$ is related to the gauge-invariant field strength $F_3$ via
\beq
G_3=F_3-C_0H
\eeq
and so the large gauge transformation corresponds to $C_0\mapsto C_0+1$, which is the generator $T$ of the S-duality group $SL(2,\Z)$.

The Bianchi identity $G_0H=0$ implies that $G_0=0$ and so there is no braneless embedding of the SU(2) WZW model in massive IIA.  Instead such an embedding will require $k$ D6-branes intersecting the 3-sphere.  In the above NS5-brane realization, this reflects the familiar fact that an NS5-brane is confined by $k$ D6's when the Romans' mass is equal to $k$.

\section{Torsion corrections from the Freed-Witten anomaly} \label{fwsec}

\subsection{The Freed-Witten anomaly}

In Ref.~\cite{FW} Freed and Witten have demonstrated that global worldsheet anomaly cancellation dictates the condition
\beq
W_3+H=dF \label{fw}
\eeq
on the worldvolume of a D$p$-brane wrapping a compact cycle $N\subset M$ in the type II string theory spacetime $M$.  Here $W_3$ is the third Stiefel-Whitney class of the normal bundle of $N$ in $M$, which vanishes if $N$ is $spin^c$ as $M$ and $N$ are both orientable in type II if the D-brane is to carry a RR charge and $M$ is $spin^c$.  More generally $W_3$ is a $\Z_2$-valued class in the third integral cohomology of $N$.  $H$ is, as usual, the NSNS 3-form field strength pulled back to $N$, or rather the associated class in integral cohomology.  $dF$ is, as a differential form, the exterior derivative of the 2-form field strength of the D-brane worldvolume's $U(1)$ gauge field, which is the magnetic monopole charge and so is Poincar\'e dual, in $N$, to boundary of a D$(p-2)$ brane that ends on our D$p$-brane.  

All three terms in Eq.~(\ref{fw}) are integral cohomology classes, thus as usual only the free part of $dF$ may be locally interpreted as the derivative of a field strength 2-form.  In this context the magnetic monopole charge $dF$ is still Poincar\'e dual to a codimension 3 submanifold of $N$ on which a lower-dimensional D-brane ends, however it may be that this submanifold is homologically trivial when included into the spacetime $M$.  An example of such a phenomenon occurs in the SU(3) WZW model and is described in Ref.~\cite{MMS}.  In this instance one finds that, instead of a D$(p-2)$-brane ending on the D$p$, a D$(p-4)$-brane ends on the D$p$.  In general the differential $d_{2j+1}$ will give us information about D$(p-2j)$-brane insertions on the D$p$-brane.  The inserted branes extend outward from the D$p$-brane until either they hit another brane on which Eq.~(\ref{fw}) permits them to end, or else until they reach the end of the spacetime.  Such configurations of branes ending on branes are referred to as baryons in Ref.~\cite{bbads}, as in some examples they represent baryonic vertices in a dual conformal field theory.

\subsection{An aside: Physical interpretation of Freed-Witten}

While in this note we are only interested in applying the Freed-Witten anomaly cancellation condition (\ref{fw}) and not in deriving it, we will now try to provide a physical interpretation of the condition.  The uninterested reader may skip to the next subsection.  Whenever there is a loop in spacetime, even if the loop is contractible, one has a choice of boundary conditions for the various spinor fields.  They may either remain invariant when they encircle the cycle, or else they may change sign.  As one deforms a loop, the spinor is transported along a $spin$ bundle, which is a choice of square root of the tangent bundle where the above sign choices are interpreted as sign choices in the square root.  However, sometimes a given set of choices is not compatible with this transport.  In fact, sometimes no set of boundary conditions is compatible.  If there exists a compatible set of fermion boundary conditions, one says that the manifold is $spin$.  Notice that a manifold, such as $\cp^2$, can be nonspin even if it is simply-connected.  This is because even the choice of boundary condition of the contractible cycle is not invariant as the contractible cycle is transported around the nontrivial 2-cycle of $\cp^2$.  

We conclude that in general a fermion partition function can only be defined if the manifold is $spin$.  If the fermion is charged under a $U(1)$ gauge symmetry then the partition function is not a section of the $spin$ bundle, but rather a section of a $spin^c$ bundle, which is the tensor product of the $spin$ bundle by the $U(1)$ gauge bundle.  In particular, even if the spacetime is not $spin$ and so the $spin$ bundle does not exist, because the product of transition functions on a three-way overlap is $-1$ instead of $+1$, there may exist a $U(1)$ gauge bundle that fails to satisfy the triple overlap condition on the same overlaps.  In this case the tensor product bundle satisfies the triple overlap condition and so the fermions exist.  Such a choice of $U(1)$ bundle is called a $spin^c$ structure, and it exists when $W_3=0$.  

We may understand this cancellation in terms of $\Z_2$-valued cohomology.  A bundle is not $spin$ if it has a second Stiefel-Whitney $w_2\neq 0\in\H^2(M;\Z_2)$ in the second cohomology group.  This obstruction is canceled by a $U(1)$ bundle if the image in $\Z_2$ cohomology of the Chern class $F\in\H^2(M;\Z_2)$ of the $U(1)$ bundle is another $\Z_2$ torsion class which precisely cancels $w_2$
\beq
w_2+F=0.
\eeq 
An honest bundle must have a Chern class valued in integral cohomology, not in cohomology with $\Z_2$ coefficients.  However using the short exact sequence
\beq
0\longrightarrow\Z\stackrel{\times2}{\longrightarrow}\Z\longrightarrow\Z_2\longrightarrow 0
\eeq
one can construct the long exact cohomology sequence
\beq
\longrightarrow \H^2(M;\Z)\longrightarrow\H^2(M;\Z_2)\stackrel{\beta}{\longrightarrow}\H^3(M;\Z)\longrightarrow \label{seq}
\eeq
where $\beta$ is referred to as a Bockstein homomorphism.  The sequence (\ref{seq}) implies that an element $w_2+F$ of $\H^2(M;Z_2)$ can be lifted to an honest Chern class, that is an element of $\H^2(M;Z)$, whenever it is in the kernel of the Bockstein.  This means that the image of the Bockstein
\beq
\beta(w_2+F)=W_3+dF \label{fwlight}
\eeq 
is the obstruction to the existence of the bundle and therefore to the existence of the fermion.  The Freed-Witten anomaly however is not precisely the condition that (\ref{fwlight}) vanishes, there is also an $H$ term. This $H$ term does not appear in QED, but is an additional complication that arises in supergravity.

In type II supergravity theories the consistency condition for spinors is weaker than in QED.  We have seen that a charged fermion in QED lives in a $spin^c$ bundle, which is the tensor product of a $spin$ bundle with a $U(1)$ bundle.  Likewise a charged fermion that is described by a fundamental string lives in the tensor product of a $spin^c$ bundle with a PU($\mathcal H$) bundle, where PU($\mathcal H$) is the projective unitary group on the Hilbert space $\mathcal H$.  While $U(1)$ bundles are classified by two-classes $F$ called Chern classes, PU($\mathcal H$) bundles are classified by 3-classes $H$ called Dixmier-Douady classes.  The total bundle on which the fermion partition function is defined is then the product of the $spin$ bundle, the U(1) bundle and the PU($\mathcal H$) bundle.  Unlike the previous case, the various characteristic classes $W_3$, $dF$ and $H$ are already integral classes, and so we do not need to worry about the obstruction to lifting them to integral classes.  However we do need to impose that $H$ cancels the other characteristic classes, which leads to the Freed-Witten anomaly (\ref{fw}).

While the above construction works for spacetimes of any dimension, for those of dimension less than 15, as in string theory, one may replace the PU($\mathcal H$) bundle with the based loop group of $E_8$.  The physical interpretations of the appearance of both of these groups is at best mysterious.  However loop groups of $E_8$ appear to be ubiquitous in string theory, for example they appear on the current algebras of the heterotic strings.  One day perhaps we will learn that these are all manifestations of the same $E_8$.  The PU($\mathcal H$) bundle, however mysterious, does provide a connection with the definition of twisted K-theory in \cite{bundlegerbes} as equivariant sections of gauge bundles over PU($\mathcal H$) bundles over spacetime.

\subsection{Quantum corrections to the Bianchi identities}

We are now ready to use the Freed-Witten anomaly to compute the torsion corrections to the BRST operator.  In particular, we need to know the quantum corrections to the constraints, which determine the action of the BRST operator on the physical fields, and we need to know the quantum corrections to the gauge transformations, which determine the action of of the BRST operator on the ghosts.

While we are searching for a classification of fluxes, the Freed-Witten anomaly is a condition on branes.  To convert a condition on branes into a condition on fluxes we use an argument based on Gauss' law that has appeared, for example, in Ref.~\cite{m2quant}.  As a differential form in de Rham cohomology, a field strength is determined by its integral over the cycles which represent various homology classes.  This notion is easily extended to the full integral cohomology by replacing integration with the homology/cohomology pairing.  Thus when we speak of a $p$-flux on a $p$-cycle, we will really be referring to the pairing of the corresponding cohomology and homology classes.  As has been argued in Ref.~\cite{MooreWitten}, the RR $p$-flux on a topologically trivial $p$-cycle measures the D$(8-p)$-brane charge that is linked by the trivial cycle.  In particular, D-brane charge in the full quantum theory is classified by integral homology, and so this pairing also includes torsion terms.  

Witten has argued in Ref.~\cite{m2quant} that even when a cycle is topologically nontrivial, the consistency conditions for the fluxes supported on the cycle, being local, cannot depend on whether at some far away place the cycle degenerates and the flux is sourced by a D-brane.  Thus the only consistent RR $p$-fluxes, even on noncontractible cycles, are supposedly those fluxes that could be sourced by a D($8-p$)-brane had the cycle been contractible.

The twisted K-theory classification only applies to fluxes in the absence of D-brane charges.  This is because D-brane charges shift the Bianchi identities and so shift the set of consistent fluxes away from twisted K-theory.  Thus the $p$-cycle on which we are measuring the $p$-flux must not intersect any D-branes.  However, if $W_3+H$ is nonzero on the worldvolume of the D$(8-p)$-brane that may source this flux, then there will be lower D-brane insertions ending on the D$(8-p)$-brane.  The other insertions may either end on other D-branes, whose total $H+W_3$ cancels that of our original brane, or they            may continue to the end of spacetime.  In the first case, the total $H+W_3$ vanishes, where this addition is defined after the various forms are pushed forward from the brane worldvolumes $N$ on to the spacetime $M$, alternately one may think of both D-branes as a single disconnected D-brane which is therefore subject to Freed-Witten.  In the second case, the semi-infinite inserted branes will intersect our $p$-cycle, no matter how distant the D($8-p$) is, and so invalidate our initial hypothesis that our spacetime in fact contains no branes.  Thus the Freed-Witten anomaly leads to quantum corrections, that is torsion corrections, to the Bianchi identity which impose that the $p$-flux on any cycle can be generated by a brane for which $W_3+H$ vanishes.

A necessary, but not sufficient, condition for the vanishing of $W_3+H$ on this D-brane worldvolume is that the $p$-flux be in the kernel of the AHSS differential
\beq
d_3=Sq^3+H\cup.
\eeq
To see this, we simplify matters by considering a spacetime which is homogenous in the radial direction from the D($8-p$), which is the direction followed by any brane insertions.  In particular if one is classifying fluxes on time slices then one may identify this radial direction with time, as in Ref.~\cite{MMS}.  On each 9-dimensional radial slice Gauss' law then ensures that the $p$-flux is Poincar\'e dual to the $(9-p)$-dimensional D($8-p$)-brane worldvolume.  The $H$ flux in the Freed-Witten anomaly (\ref{fw}) is the pullback of the $H$ flux in the bulk spacetime, thus its pushforward back to the bulk reproduces the $H$ term in $d_3$.  $Sq^3$ on the other hand is defined to be the pushforward of the third Stiefel-Whitney class, $W_3$, of the normal bundle of the cycle Poincar\'e dual to our $p$-flux $G_p$, which is precisely our D-brane worldvolume.  Thus the image of $d_3G_p$ is precisely the pushforward of $W_3+H$ under the inclusion $i:N\hookrightarrow M$, which must vanish if $W_3+H$ is indeed zero as the pushforward map is linear
\beq
d_3G_ p=Sq^3G_p+H\cup G_p=i_*(W_3+H)=0.
\eeq

\subsection{A Mod 3 Torsion Freed-Witten Anomaly}

We have now argued that a necessary condition on any $p$-flux in a brane-free world is that it be in the kernel of $d_3$.  We cannot prove that a necessary and sufficient condition is that it be in the kernel of the entire set of $d_{2j+1}$ operators.  However the results of Ref.~\cite{MMS} were consistent with the suggestion that the kernel of this collection of operators is precisely the set of fluxes dual to branes for which $W_3+H=0$.  This conjecture would be very strong, as $d_5$ for example contains the mod 3 Milnor primitive $Q_1$ \cite{Chris}, and at higher order one finds the mod 5 Milnor primitive, which are conditions on the $\Z_3$ and $\Z_5$ torsions.  This is in contrast to $W_3$, which is only a condition on the $\Z_2$ torsion, and so it is difficult to see how it may reproduce the other conditions.  One physical source of $\Z_3$ and $\Z_5$ torsion conditions in string theory is the set of $E_8$ triples, investigated in Ref.~\cite{tfs}.  A similar $\Z_3$ anomaly has appeared in M-theory \cite{FluxQuant}, where it was noted that the Chern-Simons term is at level $1/6$ and so only a conspiracy of anomaly-cancellations allows the modulo 2 and modulo 3 parts of the partition function to be well-defined.  Perhaps the $\Z_3$ torsion term is a characteristic class of a generalization of a $spin^c$ structure involving the cube root of an $E_8$ bundle that appeared in the M-theory context.

However it may be that in the low dimensions of interest to string theory such higher torsion terms do not appear.  In Ref.~\cite{MMS} the authors have demonstrated that in the SU(3) WZW model, which may be embedded in type II string theory, there is only $\Z_2$ torsion and $d_5$ successfully implements the Freed-Witten condition.  SU($N$) models at higher $N$ have higher torsions, but cannot be embedded in type II string theory so there is still no counterexample to this conjecture.  

To construct a potential counterexample we need a 10-dimensional manifold on which 
the $\Z_3$ primitive $Q_1$ is nontrivial.  One example is the Euclidean space 
$S^3/\Z_3\times S^7/\Z_3$ where we have quotiented by the $\Z_3$ subgroups of the 
free circle actions on the spheres.  This is equivalently a $T^2$ bundle over 
$\cp^1\times\cp^3$.  Such a compactification will never be Ricci flat, and so in 
practice one may wish to replace $\cp^1$ by a higher genus 
Riemann surface, but this replacement is inconsequential for a calculation 
of $Q_1$.  Acting on an integer cohomology class, $Q_1$ is just minus a Bockstein 
of the Steenrod power $P^1$. If the degree 1 and 2 cohomologies of 
$S^3/Z_3$ are generated by $a_1$ and $a_2$ and those of $S^7/\Z_3$ by $b_1$ and $b_2$ then the Bockstein $\beta$ takes $a_1$ to $-a_2$ and $b_1$ to $-b_2$.  Therefore
\beq
Q_1(a_1b_2-a_2b_1)= Q_1(a_1b_2)=-\beta(a_1 P^1 b_2)=-\beta(a_1)b_2^3=b_1 b_2^3\neq 0.
\eeq
In particular this may exclude a D6-brane Poincar\'e dual to $a_1b_2-a_2b_1$, although such a brane does not suffer from any known anomaly.  In the interpretation of Ref.~\cite{MMS} this anomaly could be canceled by a $Z_3$-charged D2 ending on a $T^2$ with one leg on each sphere, but in this compact spacetime there is no place where the other end of the D2 can terminate.  

This example should be investigated in the future.  The K-theory classification of D-branes in IIA appears to indicate that this D6-brane should not exist, and the corresponding RR 3-flux should not exist in IIB.  If this is indeed the case, it would be interesting to find a topological characterization of this obstruction and its cure.  This would provide a 3-torsion generalization of the $spin^c$ structure required by fermions on the worldvolume of a D-brane, which yields the 2-torsion constraint in the Freed-Witten anomaly.

In summary, we have shown that the Freed-Witten anomaly provides 2-torsion corrections to the Bianchi identity that imply that $p$-fluxes must indeed be in the kernel of the AHSS differential $d_3$, and we have suggested that the higher differentials may yield a necessary and sufficient condition for the vanishing of the Freed-Witten anomalies.

\subsection{Quantum corrections to the gauge transformations}

Our next goal is to try to find the quantum corrections to the gauge transformations, and to compare them with the images of the AHSS differentials.

Again we will consider a $p$-cycle $X^p$ and the $p$-flux $G_p$ that it supports.  Now we want to know not if $G_p$ is consistent, but rather we want to know whether it can be gauge transformed out of existence.  To construct a quantum gauge transformation, let us imagine that there is a D$(10-p)$ brane far away.  One may think that this is problematic as on dimensional grounds the $p$-cycle and $(11-p)$-dimensional worldvolume will intersect, but again we will restrict attention to configurations in which there is a radial symmetry about the D-brane, and so the $p$-cycle will intuitively be kept at a fixed, finite distance.  It is not necessary that there be a total D-brane charge, there could be an antibrane nearby whose charges cancel that of the original brane.  However, the presence of such a pair should not affect the available gauge symmetries in the system, as they can be spontaneously pair created and destroyed.

If our D$(10-p)$-brane wraps a cycle on which $W_3+H$ is nontrivial, then the Freed-Witten condition implies that it will contain a worldvolume magnetic source $dF$ equal to $W_3+H$.  Such a magnetic source is a lower-dimensional D-brane that ends on our D($10-p$).  In particular, we have argued in the last section that if the pushforward of $W_3+H$ onto the bulk is nonzero 
\beq
i:N\hookrightarrow M\hsp i_*:H^*(N)\longrightarrow H^*(M)\hsp i_*(W_3+H)\neq 0
\eeq
then the inserted brane will be a D($8-p$)-brane which extends radially from the D($10-p$).  As the D($8-p$)-brane is extended in the radial direction, we now need to take care that it does not intersect our $p$-cycle.  Fortunately, the intersection of the D($8-p$) with each 9-dimensional radial slice is $(8-p)$-dimensional, and so it can avoid our $p$-cycle.

We now are ready to calculate the gauge transformations of $G_p$.  These gauge transformations are the admissible transition functions of $G_p$ when one moves from patch to patch.  In particular, one may calculate the holonomy of $G_p$ as one encircles the D$(10-p)$.  Such a journey, which occurs in parameter space and not in time, sweeps out a $(p+1)$-dimensional patch $Y^{p+1}$ which is intuitively just our $p$-cycle times a circle, although the topology of the spacetime may force the topology of the $p$-cycle to change during the trip.  Mathematically, the topology of the $p$-cycle at each point is a level set in a circle-valued Morse function.  As $X$ interpolates between the $p$-cycle and itself it has no boundary, however one may use Stoke's theorem to calculate the monodromy of $X$ as the $p$-cycle sweeps out $X$, encircling the D$(10-p)$.  

We will write this monodromy as an integral, although we are really considering a homology/cohomology pairing
\beq
\Delta \int_{X^p} G_p=\int_{Y^{p+1}}dG_p. \label{x}
\eeq
We now remember that D($8-p$)-brane charge is Poincar\'e dual to the source $dG_p$, which is not exact as $G_p$ is not gauge invariant and so not globally defined.  Thus the right hand side of (\ref{x}) is the intersection number of the trajectory $Y^{p+1}$ and the D(8-p)-brane worldvolume.  Eq.~(\ref{x}) then implies that the transformation of the $p$-flux $G_p$ is the Poincar\'e dual of the D$(8-p)$-brane worldvolume, which we have seen may be obtained from the original D$(10-p)$-brane using the Freed-Witten anomaly
\beq
\Delta \int_{X^p} G_p=i_*(W_3+H).
\eeq
We may now again use the above definition of the Steenrod squares to write the pushforward of $W_3$ as $Sq^3$ of the Poincar\'e dual $\eta$ of the D($10-p$)-brane worldvolume
\beq
\Delta \int_{X^p} G_p=i_*(W_3+H)=(Sq^3+H\cup)x=d_3 x.
\eeq
Thus we find that quantum gauge transformations allow $G_p$ to vary by any class in the image of $d_3$, identifying the quotient in the first step of the AHSS as the identification of gauge orbits under gauge transformations that are not annihilated by the pushforward.  One may again, following Ref.~\cite{MMS}, identify the higher differentials with the gauge transformations that are killed by the pushforward operator but are captured by various secondary cohomology operations.

\section{Summary}

We have argued that in type II supergravity theories the set of RR field strengths that satisfy the quantum-corrected Bianchi identities quotiented by the quantum-corrected large gauge transformations is the twisted K-theory of the spacetime, more precisely, twisted $\K^0$ for IIA and $\K^1$ for IIB.  In particular, we have identified the Atiyah-Hirzebruch spectral sequence (AHSS) construction of twisted K-theory as the usual BRST formulation in quantum field theory, albeit for discrete transformations so the ghosts have no propagating degrees of freedom.  The differential operators of the AHSS have been identified with BRST operators for these large gauge transformations.

In the classical limit, the large gauge transformations are just the shifts of the RR gauge connection that preserve the Wilson loops, which form an integral lattice of the de Rham cohomology.  The corresponding constraints arise classically from the Bianchi identity.  Quantum corrections to the Bianchi identity arise by supposing that the field strength is sourced by a D-brane and considering the Freed-Witten anomaly on the D-brane's worldvolume.  Quantum corrections to the gauge transformations similarly come from considering the holonomy of a flux as one encircles a D-brane which is afflicted with a Freed-Witten anomaly.

We hope that this construction of the twisted K-theory classification entirely within the usual framework of quantum field theory will be useful for attempted constructions of K-theory based quantum field theories, as one sees that no new technology needs to be introduced.  In particular, here we have classified only the field strengths, but if one is also interested in the gauge connections then the BRST cohomology will yield a differential twisted K-theory.  One may then compare this form of differential twisted K-theory with that which has been conjectured to exist in string theory in Ref.~\cite{Freed}, thereby testing the conjecture.

One may also apply this strategy to the longstanding problem of reconciling the twisted K-theory classification with S-duality.  To do this, one needs to also consider the Bianchi identities of the NSNS fields and the NSNS gauge transformations.  To consider these in parallel with those of the RR fields may require the BV formalism.

\section* {Acknowledgement}

I would like to thank Allan Adams, Glenn Barnich, Nazim Bouatta, Christopher Douglas, Michael Douglas, Marc Henneaux, Mike Hill, Stanislav Kuperstein and even Daniel Persson for speaking with me.  I would like to thank the organizers of Problemi Attuali di Fisica Teorica at Vietri sul Mare for inviting me to give this talk, as it motivated me to finally write this up after four years. 

My work is partially supported by IISN - Belgium (convention 4.4505.86), by the ``Interuniversity Attraction Poles Programme -- Belgian Science Policy'' and by the European Commission programme MRTN-CT-2004-005104, in which I am associated to V. U. Brussel.

\end{document}

\bibitem{}
,
{\it},
[{\tt arXiv:hep-th/}].

\bibitem{Manjarin}
J.~J.~Manjarin,
{\it Topics on D-brane Charges with B-fields},
[{\tt arXiv:hep-th/0405074}].

\bibitem{Baryons}
E.~Witten,
{\it Baryons and Branes in Anti de Sitter Space},
[{\tt arXiv:hep-th/9805112}].

\bibitem{MMS}
J.~Maldacena, G.~Moore and N.~Seiberg,
{\it D-Brane Instantons and K-Theory Charges},
[{\tt arXiv:hep-th/0108100}].

\bibitem{Feb}
J.~Evslin,
{\it Twisted K-Theory from Monodromies},
[{\tt arXiv:hep-th/0302081}].

\bibitem{KS}
I.~R.~Klebanov and M.~J.~Strassler,
{\it Supergravity and a Confining Gauge Theory: Duality Cascades and $\chi$SB-Resolution of Naked Singularities},
[{\tt arXiv:hep-th/0007191}].

\bibitem{Nov}
J.~Evslin,
{\it IIB Soliton Spectra with All Fluxes Actived},
[{\tt arXiv:hep-th/0211172}].

\bibitem{Ftheory}
C.~Vafa,
{\it Evidence for F-Theory},
[{\tt arXiv:hep-th/9602022}].

\bibitem{Oscar}
O.~Loaiza-Brito,
{\it Instantonic Branes, Atiyah-Hirzebruch Spectral Sequence, and SL(2,Z) Duality of N=4 SYM},
[{\tt arXiv:hep-th/0311028}].

\bibitem{KPV}
S.~Kachru, J.~Pearson and H.~Verlinde,
{\it Brane/Flux Annihilation and the String Dual of a Non-Supersymmetric Field Theory},
[{\tt arXiv:hep-th/0112197}].

\bibitem{BCMMS}
P. Bouwknegt, A. Carey, V. Mathai, M. Murray and D. Stevenson,
{\it Twisted K-theory and K-theory of bundle gerbes},
Comm. Math. Phys. {\bf 228} (2002) 17-45,
[{\tt arXiv:hep-th/0106194}].

\bibitem{MathStev}
V.~Mathai and D.~Stevenson,
{\it On a Generalized Connes-Hochschild-Kostant-Rosenberg Theorem},
[{\tt arXiv:hep-th/0404329}].

\bibitem{Marolf}
D.~Marolf,
{\it Chern-Simons Terms and the Three Notions of Charge},
[{\tt arXiv:hep-th/0006117}].

\bibitem{Wati}
W.~ Taylor
{\it D-Branes in B Fields},
[{\tt arXiv:hep-th/0004141}].

\bibitem{Kapustin}
A.~Kapustin,
{\it D-Branes in a Topologically Nontrivial B-Field},
[{\tt arXiv:hep-th/9909089}].


\bibitem{PetrIIA}
P.~ Ho$\check{\textup{r}}$ava,
{\it Type IIA D-Branes, K-Theory and Matrix Theory},
[{\tt arXiv:hep-th/9812135}].

\bibitem{MN}
J.~M.~Maldacena and C.~Nunez,
{\it Towards the large N limit of pure N=1 super Yang Mills},
[{\tt arXiv:hep-th/0008001}].

\bibitem{HanWit}
A.~ Hanany and E.~Witten,
{\it Type \twob\ Superstrings, BPS Monopoles and Three-Dimensional Gauge Dynamics},
[{\tt arXiv:hep-th/9611230}].

\bibitem{BDS}
C.~Bachas, M.~Douglas and C.~Schweigert,
{\it Flux Stabilization of D-Branes},
[{\tt arXiv:hep-th/0003037}].
 
\bibitem{WittenMQCD}
E.~Witten,
{\it Solutions of Four-Dimensional Field Theories Via M-Theory},
[{\tt arXiv:hep-th/9703166}].

\bibitem{Hitoshi}
J.~Evslin, H.~Murayama, U.~Varadarajan and J.~.E.~Wang,
{\it Dial M for Flavor Symmetry Breaking},
[{\tt arXiv:hep-th/0107072}].

\bibitem{SeibergDuality}
N.~Seiberg,
{\it Electric-Magnetic Duality in Supersymmetric Non-Abelian Gauge Theories},
[{\tt arXiv:hep-th/9411149}].

\bibitem{Ken}
G.~Carlino, K.~Konishi and H.~Murayama,
{\it Dynamics of Supersymmetric $SU(n_c)$ and $USp(2n_c)$ Gauge Theories},
[{\tt arXiv:hep-th/0001036}].

\bibitem{OT}
K.~Oh and R.~Tatar,
{\it Duality and Confinement in N=1 Supersymmetric Theories from Geometric Transitions},
[{\tt arXiv:hep-th/0112040}].

\bibitem{AtiyahWitten}
M.~Atiyah and E.~Witten,
{\it M-Theory Dynamics on a Manifold of $G_2$ Holonomy},
[{\tt arXiv:hep-th/0107177}].

\bibitem{AD}
P.~C.~Argyres and M.~R.~Douglas,
{\it New Phenomena in SU(3) Supersymmetric Gauge Theory},
[{\tt arXiv:hep-th/9505062}].

\bibitem{KN}
I.~R.~Klebanov and N.~Nekrasov,
{\it Gravity Duals of Fractional Branes and Logarithmic RG Flow},
[{\tt arXiv:hep-th/9911096}].

\bibitem{BT}
J.~D.~Brown and C.~Teitelboim,
{\it Neutralization of the Cosmological Constant by Membrane Creation},
Nucl. Phys. {\bf B297}, 787, (1988).

\bibitem{bion}
N.~R.~Constable, R.~C.~Myers and O.~Tafjord,
{\it The Noncommutative Bion Core},
[{\tt arXiv:hep-th/9911136}].

\bibitem{Gibbons}
G.~W.~Gibbons,
{\it Branes as BIons},
[{\tt arXiv:hep-th/9803203}].

\bibitem{MW}
G.~Moore and E.~Witten,
{\it Self duality, Ramond-Ramond fields, and K-theory},
J. High Energy Phys. {\bf 05} (2000) 032,
[{\tt arXiv:hep-th/9912279}].

\bibitem{Townsend}
M. B. Green, C. M. Hull and P. K. Townsend,
{\it D-brane Wess-Zumino Actions, T-duality and the Cosmological Constant},
[{\tt arXiv:hep-th/9604119}].

\bibitem{VanProeyen}
E. Bergshoeff, R. Kallosh, T. Ortin, D. Roest and A. Van Proeyen,
{\it New Formulations of D=10 Supersymmetry and D8-O8 Domain Walls},
[{\tt arXiv:hep-th/0103233}].

\end{thebibliography}
\end{document}

\bibitem{}
,
{\it},
[{\tt arXiv:hep-th/}].

\bibitem{}
,
{\it},
[{\tt arXiv:hep-th/}].

\bibitem{NdW}
B. de Wit and H. Nicolai, Phys.\ Lett.\ {\bf 155B}, 47 (1985); Nucl. Phys. \textbf{B274}, 363 (1986).

\bibitem{Nic}
H. Nicolai, Phys.\ Lett.\ {\bf 187B}, 363 (1987). 

\bibitem{HW}
P.~ Ho$\check{\textup{r}}$ava and E.~ Witten,
{\sl Heterotic and Type I String Dynamics from Eleven Dimensions},
Nucl. Phys. {\bf B460}, 506, (1996), 
[{\tt arXiv:hep-th/9510209}].

\bibitem{FH}
M. Fabinger and P.~ Ho$\check{\textup{r}}$ava
{\it Casimir Effect between World-Branes in Heterotic M-Theory},
Nucl Phys. {\bf B580}, 243, (2000), 
[{\tt arXiv:hep-th/0002073}].

\bibitem{BEM}
P. Bouwknegt, J.~Evslin, and V. Mathai, 
{\it T-duality: Topology Change from H flux}, 
[{\tt arXiv:hep-th/0306062}].

\bibitem{FluxQuant}
E.~ Witten, {\sl On Flux Quantization in M-Theory and the Effective 
Action},
J. Geom. Phys. {\bf 22},1 , (1997), 
[{\tt arXiv:hep-th/9609122}].

\bibitem{GM}
C.~ Gomez, J.~ J.~ Manjarin,
{\it Dyons, K-theory and M-theory}, 
[{\tt arXiv:hep-th/0111169}].

\bibitem{allan}
A.~Adams and J.~Evslin, {\it The Loop Group of $E_8$ and K-Theory from $11d$},
JHEP {\bf 02}, 29 (2003), 
[{\tt arXiv:hep-th/0203218}]. 

\bibitem{Morrison}
D.~R.~Morrison, 
{\it "Half $K3$ surfaces"}
talk at Strings 2002, Cambridge,
\\
http://www.damtp.cam.ac.uk/strings02/avt/morrison/

\bibitem{MS}
G.~Moore and N.~Saulina,
{\it T-duality, and the K-theoretic partition function of Type \twoa
superstring theory},
[{\tt arXiv:hep-th/0206092}].

\bibitem{Kentaro}
K.~Hori
{\sl Consistency Conditions for Five-Brane Theory in M-Theory on $R^5/\Z_2$ Orbifold},
Nucl. Phys. {\bf B539}, 35, (1999), hep-th/9805141.

\bibitem{slow}
A.~ Adams, J.~ Evslin, and U.~ Varadarajan,
(To Appear).

\bibitem{Stong}
R.~Stong,
{\it Calculation of $\Omega_{11}^{\textup{spin}}({\mathbf{K}(\Z,4))})$},
in Unified String Theories, eds. M.~B.~Green and D.~J.~Gross, World 
Scientific, 1986.

\bibitem{Horava}
P.~Ho\v rava, 
(Unpublished).

\bibitem{Hull}
C.~ M.~ Hull,
{\it Massive String Theories from M-Theory and F-Theory},
JHEP {\bf 9811} (1998) 027, 
[{\tt arXiv:hep-th/9811021}].

\bibitem{Sethi}
S.~ Sethi,
(Unpublished).

\bibitem{Manjarin}
C.~ Gomez and J. J. Manjarin,
(To Appear).

\bibitem{KentaroIndexTDuality}
K.~Hori, {\it D-branes, T-duality, and index theory},
Adv. Theor. Math. Phys. {\bf 3} (1999) 281-342,
[{\tt arXiv:hep-th/9902102}].

\bibitem{AABL}
E.~\'Alvarez, L.~\'Alvarez-Gaum\'e, J.L.F.~Barb\'on and Y.~Lozano,
{\it Some global aspects of duality in string theory},
Nucl. Phys. {\bf B415} (1994) 71-100,
[{\tt arXiv:hep-th/9309039}].

\bibitem{RV}
M.~Ro\v cek and E.~Verlinde,
{\it Duality, quotients, and currents},
Nucl. Phys. {\bf 373} (1992) 630-646,
[{\tt arXiv:hep-th/9110053}].

\end{thebibliography}
\end{document}